\documentclass{emulateapj}

\newcommand{\npatchtext}{four }

\newcommand{\ndiv}{4~}
\newcommand{\ndivtext}{four }

\newcommand{\nhat}{\hat{\mathbf{n}}}
\newcommand{\bl}{{\bm \ell}}
\newcommand{\ave}[1]{\left\langle#1\right\rangle}

\newcommand{\LCDM}{$\Lambda$CDM }

\newcommand{\lbegin}{500}
\newcommand{\lend}{10000}

\newcommand{\be}{\begin{equation}}
\newcommand{\ee}{\end{equation}}
\newcommand{\ba}{\begin{eqnarray}}
\newcommand{\ea}{\end{eqnarray}}
\newcommand{\ara}{148}
\newcommand{\arb}{218}
\newcommand{\aroneDust}{150 GHz}
\newcommand{\artwoDust}{220 GHz}
\newcommand\pmerrors[2]{^{+#1}_{-#2}}

\usepackage[citebordercolor={0 .5 .5}]{hyperref}

\usepackage{graphicx}        
\usepackage{bm}              
\usepackage{natbib}
\usepackage{mathrsfs}

\newcommand{\arone}{148\,GHz}
\newcommand{\artwo}{218\,GHz}
\newcommand{\arthree}{277\,GHz}

\newcommand{\commentx}[1]{}

\newcommand{\etal}{et al.\,}  

\newcommand{\ra}[3]   
   {\makebox[1.5em][r]{#1}\makebox[1.5em][r]{#2} \makebox[2em][r]{#3}}

\newcommand{\hms}[3]  
   {${#1}^{\mathrm{h}}{#2}^{\mathrm{m}}{#3}^{\mathrm{s}}$}

\newcommand{\hmin}[2]  
   {\ensuremath{{#1}^{\mathrm{h}}{#2}^{\mathrm{m}}}}
   
\newcommand{\hours}[1]  
   {\ensuremath{{#1}^{\mathrm{h}}}}

\newcommand{\dms}[3]  
   {\ensuremath{{#1}\degree{#2}\arcminute{#3}\arcsecond}}

\newcommand{\dm}[2]  
   {\ensuremath{{#1}\degree{#2}\arcminute}}

\newcommand{\ukcmb}  
           {\ensuremath{\micro \kelvin_\mathrm{cmb}}}

\newcommand{\uk}  
           {\ensuremath{\micro \kelvin}}

\newcommand{\fdeg} 
           {\hbox{$.\!\!^{\circ}$}}

\usepackage{color}

\usepackage[mediumspace,Gray,squaren]{SIunits}
\usepackage{rotating}
\usepackage{graphicx}
\shorttitle{ACT 2008 Power Spectrum}
\shortauthors{S. Das \etal}

\begin{document}

\title{The Atacama Cosmology Telescope:  A Measurement of
  the Cosmic Microwave Background Power Spectrum at 148 and 218 GHz from the 2008  Southern Survey}


\author{
Sudeep~Das\altaffilmark{1,2,3},
Tobias~A.~Marriage\altaffilmark{3,4},
Peter~A.~R.~Ade\altaffilmark{5},
Paula~Aguirre\altaffilmark{6},
Mandana~Amiri\altaffilmark{7},
John~W.~Appel\altaffilmark{2},
L.~Felipe~Barrientos\altaffilmark{6},
Elia~S.~Battistelli\altaffilmark{8,7},
J.~Richard~Bond\altaffilmark{9},
Ben~Brown\altaffilmark{10},
Bryce~Burger\altaffilmark{7},
Jay~Chervenak\altaffilmark{11},
Mark~J.~Devlin\altaffilmark{12},
Simon~R.~Dicker\altaffilmark{12},
W.~Bertrand~Doriese\altaffilmark{13},
Joanna~Dunkley\altaffilmark{14,2,3},
Rolando~D\"{u}nner\altaffilmark{6},
Thomas~Essinger-Hileman\altaffilmark{2},
Ryan~P.~Fisher\altaffilmark{2},
Joseph~W.~Fowler\altaffilmark{13,2},
Amir~Hajian\altaffilmark{9,3,2},
Mark~Halpern\altaffilmark{7},
Matthew~Hasselfield\altaffilmark{7},
Carlos~Hern\'andez-Monteagudo\altaffilmark{15},
Gene~C.~Hilton\altaffilmark{13},
Matt~Hilton\altaffilmark{16,17},
Adam~D.~Hincks\altaffilmark{2},
Ren\'ee~Hlozek\altaffilmark{14},
Kevin~M.~Huffenberger\altaffilmark{18},
David~H.~Hughes\altaffilmark{19},
John~P.~Hughes\altaffilmark{20},
Leopoldo~Infante\altaffilmark{6},
Kent~D.~Irwin\altaffilmark{13},
Jean~Baptiste~Juin\altaffilmark{6},
Madhuri~Kaul\altaffilmark{12},
Jeff~Klein\altaffilmark{12},
Arthur~Kosowsky\altaffilmark{10},
Judy~M~Lau\altaffilmark{21,22,2},
Michele~Limon\altaffilmark{23,12,2},
Yen-Ting~Lin\altaffilmark{24,3,6},
Robert~H.~Lupton\altaffilmark{3},
Danica~Marsden\altaffilmark{12},
Krista~Martocci\altaffilmark{25,2},
Phil~Mauskopf\altaffilmark{5},
Felipe~Menanteau\altaffilmark{20},
Kavilan~Moodley\altaffilmark{16,17},
Harvey~Moseley\altaffilmark{11},
Calvin~B.~Netterfield\altaffilmark{26},
Michael~D.~Niemack\altaffilmark{13,2},
Michael~R.~Nolta\altaffilmark{9},
Lyman~A.~Page\altaffilmark{2},
Lucas~Parker\altaffilmark{2},
Bruce~Partridge\altaffilmark{27},
Beth~Reid\altaffilmark{28,2},
Neelima~Sehgal\altaffilmark{21},
Blake~D.~Sherwin\altaffilmark{2},
Jon~Sievers\altaffilmark{9},
David~N.~Spergel\altaffilmark{3},
Suzanne~T.~Staggs\altaffilmark{2},
Daniel~S.~Swetz\altaffilmark{12,13},
Eric~R.~Switzer\altaffilmark{25,2},
Robert~Thornton\altaffilmark{12,29},
Hy~Trac\altaffilmark{30,31},
Carole~Tucker\altaffilmark{5},
Ryan~Warne\altaffilmark{16},
Ed~Wollack\altaffilmark{11},
Yue~Zhao\altaffilmark{2}
}
\altaffiltext{1}{Berkeley Center for Cosmological Physics, LBL and
Department of Physics, University of California, Berkeley, CA, USA 94720}
\altaffiltext{2}{Joseph Henry Laboratories of Physics, Jadwin Hall,
Princeton University, Princeton, NJ, USA 08544}
\altaffiltext{3}{Department of Astrophysical Sciences, Peyton Hall, 
Princeton University, Princeton, NJ USA 08544}
\altaffiltext{4}{Current address: Dept. of Physics and Astronomy, The Johns Hopkins University, 3400 N. Charles St., Baltimore, MD 21218-2686}
\altaffiltext{5}{School of Physics and Astronomy, Cardiff University, The Parade, 
Cardiff, Wales, UK CF24 3AA}
\altaffiltext{6}{Departamento de Astronom{\'{i}}a y Astrof{\'{i}}sica, 
Facultad de F{\'{i}}sica, Pontific\'{i}a Universidad Cat\'{o}lica,
Casilla 306, Santiago 22, Chile}
\altaffiltext{7}{Department of Physics and Astronomy, University of
British Columbia, Vancouver, BC, Canada V6T 1Z4}
\altaffiltext{8}{Department of Physics, University of Rome ``La Sapienza'', 
Piazzale Aldo Moro 5, I-00185 Rome, Italy}
\altaffiltext{9}{Canadian Institute for Theoretical Astrophysics, University of
Toronto, Toronto, ON, Canada M5S 3H8}
\altaffiltext{10}{Department of Physics and Astronomy, University of Pittsburgh, 
Pittsburgh, PA, USA 15260}
\altaffiltext{11}{Code 553/665, NASA/Goddard Space Flight Center,
Greenbelt, MD, USA 20771}
\altaffiltext{12}{Department of Physics and Astronomy, University of
Pennsylvania, 209 South 33rd Street, Philadelphia, PA, USA 19104}
\altaffiltext{13}{NIST Quantum Devices Group, 325
Broadway Mailcode 817.03, Boulder, CO, USA 80305}
\altaffiltext{14}{Department of Astrophysics, Oxford University, Oxford, 
UK OX1 3RH}
\altaffiltext{15}{Max Planck Institut f\"ur Astrophysik, Postfach 1317, 
D-85741 Garching bei M\"unchen, Germany}
\altaffiltext{16}{Astrophysics and Cosmology Research Unit, School of
Mathematical Sciences, University of KwaZulu-Natal, Durban, 4041,
South Africa}
\altaffiltext{17}{Centre for High Performance Computing, CSIR Campus, 15 Lower
Hope St., Rosebank, Cape Town, South Africa}
\altaffiltext{18}{Department of Physics, University of Miami, Coral Gables, 
FL, USA 33124}
\altaffiltext{19}{Instituto Nacional de Astrof\'isica, \'Optica y 
Electr\'onica (INAOE), Tonantzintla, Puebla, Mexico}
\altaffiltext{20}{Department of Physics and Astronomy, Rutgers, 
The State University of New Jersey, Piscataway, NJ USA 08854-8019}
\altaffiltext{21}{Kavli Institute for Particle Astrophysics and Cosmology, Stanford
University, Stanford, CA, USA 94305-4085}
\altaffiltext{22}{Department of Physics, Stanford University, Stanford, CA, 
USA 94305-4085}
\altaffiltext{23}{Columbia Astrophysics Laboratory, 550 W. 120th St. Mail Code 5247,
New York, NY USA 10027}
\altaffiltext{24}{Institute for the Physics and Mathematics of the Universe, 
The University of Tokyo, Kashiwa, Chiba 277-8568, Japan}
\altaffiltext{25}{Kavli Institute for Cosmological Physics, 
5620 South Ellis Ave., Chicago, IL, USA 60637}
\altaffiltext{26}{Department of Physics, University of Toronto, 
60 St. George Street, Toronto, ON, Canada M5S 1A7}
\altaffiltext{27}{Department of Physics and Astronomy, Haverford College,
Haverford, PA, USA 19041}
\altaffiltext{28}{Institut de Ciencies del Cosmos (ICC), University of
Barcelona, Barcelona 08028, Spain}
\altaffiltext{29}{Department of Physics , West Chester University 
of Pennsylvania, West Chester, PA, USA 19383}
\altaffiltext{30}{Department of Physics, Carnegie Mellon University, Pittsburgh, PA 15213}
\altaffiltext{31}{Harvard-Smithsonian Center for Astrophysics, 
Harvard University, Cambridge, MA, USA 02138}

\begin{abstract}
We present measurements of the cosmic microwave background (CMB) power spectrum made
 by the Atacama Cosmology Telescope at \arone{} and \artwo{}, as well as the cross-frequency spectrum
between the two channels. Our results clearly show  the second through the seventh acoustic peaks in the CMB power spectrum.  The measurements 
of these higher-order peaks provide an additional test of  the \LCDM cosmological model.
At $\ell>3000$, we detect power in excess of the primary anisotropy spectrum of the CMB.
 At lower multipoles $500 < \ell < 3000$,  we find evidence for gravitational lensing of  the CMB in the power spectrum at the $2.8\sigma$ level. 
 We also detect a low level of Galactic dust in our maps,  which demonstrates that we can recover known faint, diffuse signals.
\end{abstract}


\setcounter{footnote}{0}

\section{INTRODUCTION}

\defcitealias{fowler/etal:prep}{F10}
\defcitealias{Hincks/etal:prep}{H09}
Accurate measurements of the arcminute scale temperature anisotropies in the mm-wave sky
are uncovering a complex, yet revealing picture \citep[][hereafter F10]{lueker/etal:2010,fowler/etal:prep}.
On intermediate scales  ($ 500 \lesssim \ell \lesssim 3000$),  the 
primordial acoustic features imprinted on the cosmic microwave background (CMB) at last scattering ($z \simeq 1100$) dominate, with  subtle 
distortions expected from  gravitational lensing by intervening large-scale structure.  This  intermediate range of multipoles is often 
called the \emph{damping tail} of the CMB, as the acoustic oscillations are exponentially damped due to photon 
diffusion \citep{silk:1968,bond/efstathiou:1984}.
On the smallest scales ($\ell \gtrsim 3000$), 
 emission from radio and dust-enshrouded star-forming galaxies, together with  the 
Sunyaev-Zel'dovich (SZ) effect \citep[][]{zeldovich/sunyaev:1969}
 dominates over primary CMB fluctuations. 
 \par
The  damping tail measurements are an additional test of the predictions of the $\Lambda$CDM cosmological model,
 a model that is an excellent fit to current CMB data \citep[e.g.,][]{larson/etal:prep,reichardt/etal:2009,brown/etal:2009},
large-scale structure measurements \citep[e.g.,][]{reid/etal:2010,percival/etal:2010}, supernova  observations
\citep[e.g.,][]{kessler/etal:2009,amanullah/etal:2010}, and a host of other astronomical observations 
(see, e.g., \citealt{spergel/etal:2007} and \citealt{komatsu/etal:prep} for reviews).  
The amplitude of the fluctuations in the damping tail is a sensitive probe of matter fluctuations at $k \simeq 0.1 - 0.25$ Mpc$^{-1}$ 
--- thus, precision measurements constrain the spectral index of the primordial curvature perturbations, $n_s$, and its variation 
with scale.  Because the positions of the high order acoustic peaks are sensitive  to the evolution of the sound speed of the universe 
and its composition, these measurements constrain the primordial helium fraction and the number of relativistic species including neutrinos \citep[see 
e.g.][]{white:2001, bashinsky/seljak:2004, trotta/hansen:2004, ichikawa/sekiguchi/takahashi:2008a,
ichikawa/sekiguchi/takahashi:2008b, komatsu/etal:prep}.  While the damping tail gives us leverage on 
early universe cosmology, the \emph{composite tail} ($\ell \gtrsim 3000$)  is sensitive to a variety of astrophysical 
phenomena, including  the  properties of the intracluster medium, the redshift distribution and clustering of dusty sub-mm galaxies 
\citep{hall/etal:2010}, and the physics of reionization \citep{huffenberger/seljak:2005,sehgal/etal:2007}.
\citet{sehgal/etal:2010} describe theoretical expectations for these small-scale measurements.    \par
In this paper, we present a measurement of the CMB power spectrum over the range of multipoles $\lbegin<\ell<
\lend$ from the Atacama Cosmology Telescope (ACT) 
 using the \arone{} and \artwo{} channel data from the 2008 observing season.
  ACT is a mm-wave, arcminute-resolution telescope \citep{fowler/etal:2007,swetz/etal:prep} custom built to make precise observations of the 
microwave sky over the damping and composite tail regimes. In recent years,  several groups have reported 
rapidly improving measurements of fluctuations over various portions of this multipole range: 
Bolocam \citep{sayers/etal:2009}, QUaD \citep{brown/etal:2009, friedman/etal:2009}, APEX-SZ
  \citep{reichardt/etal:2009a},  ACBAR \citep{reichardt/etal:2009}, SZA
  \citep{sharp/etal:2010}, BIMA \citep{dawson/etal:2006}, CBI
  \citep{sievers/etal:prep}, SPT \citep{lueker/etal:2010,hall/etal:2010} and ACT \citepalias{fowler/etal:prep}.
The current paper enhances the results of \citetalias{fowler/etal:prep} in a few ways.
First, we augment our \arone\ data with results from our \artwo\ channel. Second, the
power spectrum estimation 
methods have been revised to provide enough angular frequency resolution to detect the acoustic features on the 
damping tail.  Third, the area used for the power spectrum  analysis has been increased from $\simeq$220 deg$^2$ to $\simeq 300$ deg$^2$.
\par 
\defcitealias{marriage/etal:prepa}{M10}
This paper is one in a set of  papers describing ACT and its 2008 Southern survey.
The ACT instrument is described in \citet{swetz/etal:prep}, a \arone{} point source catalog is presented  in \citet{marriage/etal:prepa} 
(hereafter M10), and a \arone{} SZ cluster catalog is presented in \citet{marriage/etal:prepb}. \citet{menanteau/etal:prep} discuss 
the multi-wavelength followup 
of ACT clusters, while \citet{sehgal/etal:prep} present $\sigma_8$ constraints from SZ cluster detections. 
On the power spectrum side, \citet{hajian/etal:prep} report on the calibration of ACT maps 
using cross-correlations with WMAP seven-year maps, and \citet{dunkley/etal:prep} present the constraints on
cosmological parameters derived from the power spectrum presented here. 
\par
The paper is organized as follows. In Section~\ref{sec: instrument_and_obs}, we briefly review the ACT instrument
and observations made so far, touching upon mapmaking and  beam determination techniques, and the calibration of our data. 
 In Section~\ref{sec: ps_method}, we describe the method used in  power spectrum estimation. Simulations used to 
test our pipeline are discussed in Section~\ref{sec: simulations}. We present our results in Section~\ref{sec: results} 
 and discuss tests performed to 
 validate our  results.
The main sources of foreground contamination and the methods used to treat them are discussed in Section~\ref{sec:fg}.  
We discuss the lensing contribution to the power spectrum in Section~\ref{sec: cmbLensing}, before concluding in Section~\ref{sec: conclusions}.
\setcounter{footnote}{0}
\section{OBSERVATIONS AND MAPMAKING}
\label{sec: instrument_and_obs}
ACT is a 6-meter off-axis Gregorian telescope \citep{fowler/etal:2007} situated at an elevation of 5190 meters on Cerro Toco in the Atacama 
desert in northern Chile.  ACT has three frequency bands centered at 
\arone{} (2.0\,mm), \artwo{} (1.4\,mm) and \arthree{}  (1.1\,mm) with angular resolutions of roughly 1\farcm 4, 
1\farcm 0 and 0\farcm 9, respectively. The high altitude site in the arid desert is excellent for microwave observations due to  low 
precipitable water vapor and stability of the atmosphere. The tropical location of ACT  permits observations on both  
the northern and southern celestial hemispheres. Further details on the instrument  are presented in 
\citet{swetz/etal:prep}, \citetalias{fowler/etal:prep} and references therein\footnote{ACT Collaboration papers are archived at \url{http://www.physics.princeton.edu/act/}}.
As of this writing, ACT has completed three observing seasons (2007, 2008, 2009) surveying two stripes on the sky: a 
9\degree-wide stripe centered on declination -53\fdeg 5 and a $5\degree$-wide stripe centered on the celestial 
equator. The power spectrum presented here is derived from the southern stripe data from the two lower frequency 
channels in  2008. 
An important aspect of ACT's scanning strategy is the cross-linking of observations. Every point in the survey area is scanned 
along two different directions during the night. We use constant elevation scans centered on two different azimuths ---
 once when the survey area is rising on the sky, and again when it is setting 
toward the end of the observing night. In principle, this cross-linking allows an unbiased reconstruction of {\emph{all}} modes in the map,
and permits recovery of the  power spectrum for multipoles as low as a few hundred with errors 
completely dominated by cosmic variance.  
  A more detailed account of the observing strategy of ACT can be found in \citetalias{fowler/etal:prep} and 
\citetalias{marriage/etal:prepa}.
%
%
%
\subsection{Mapmaking \label{sec: mapmaking}}
For the \arone{} data, we use the maps from
\citetalias{fowler/etal:prep}, in which details of the data reduction
and mapping can be found.  We model the data as $\bm d = {\rm P}{\bm
m}+\bm n$, where $\rm{P}$ projects the  map $\bm m$ into
the time stream, and $\bm n$ is the noise, with covariance matrix
$\rm{N}$.  There are no constraints on the contents of ${\rm P}$ and
${\bm m}$, which can contain multiple components.  For ACT, in
addition to solving for the map of the sky, we solve simultaneously
for noise correlated amongst the detectors.  Under this model, the
data can be  described by 
\begin{equation}
\bm d = {\rm P}{\bm m}+ {\rm A}{\bm c} +\bm n
\label{eq:genmap}
\end{equation}
 where ${\rm A}$ are (assumed constant) patterns of correlation
across the array (such as a common mode) and ${\bm c}$ are the
timestreams associated with each pattern in ${\rm A}$.  This is
mathematically equivalent to having separate blocks in a generalized
projection matrix and a generalized map solution.  The maximum
likelihood solution, ${\tilde{\bm m}}$, is then given by solving the
standard mapping equation:
\begin{equation} 
{ {\rm P}^{\rm T} {\rm N}^{-1}{\rm P} {\tilde{\bm{m}}}={\rm P}^{\rm T}{\rm N}^{-1} \bm d.}
\label{eq:mapeq} 
\end{equation}
We solve for ${\tilde{\bm m}}$ iteratively using a preconditioned conjugate gradient (PCG) method \citep{press/teukolsky/vetterling:NRC:3e, hinshaw/etal:2007}.  

\par 

The
higher atmospheric noise in the \artwo{} data required some changes 
to the mapping pipeline.  For the \arone\ data, we found that
taking the array patterns ${\rm A}$ to be the eigenvectors corresponding to the 10 largest eigenvalues of
the data covariance matrix for each 15-minute chunk of time-ordered data (TOD) worked
well for correlated noise rejection.  For the \artwo\ data,
substantial atmospheric power remains with this prescription, and so
we altered it as follows:  first, for each TOD we take the band-limited data between
0.25 and 4 Hz, find the eigenvalues and eigenvectors of the
corresponding covariance matrix, and keep all modes with eigenvalues
larger than $3.5^2$ times the median eigenvalue (equivalently modes 
with timestream amplitudes larger than 3.5 times the median). 
 We typically find 30--50 modes.  Then, we create the
covariance matrix from the data high-pass filtered above 4 Hz, project out
the modes already found in the 0.25--4 Hz band, and keep all remaining modes with
eigenvalue larger than $2.5^2$ times the median.  We typically find
1--2 additional modes in this step. Of the several different mode removal schemes we tried,
 we found this fairly aggressive one gave the best signal-to-noise on intermediate and small 
 angular scales, where the \artwo\ data are most valuable, at the price of worse signal-to-noise 
 and slower convergence of the mapper on large scales.
Since the method estimates both the correlated modes and the map of the sky simultaneously, mode removal 
does not bias the maps, although it makes some sky map modes noisier.

The second change in the mapping is the use of a preconditioner with
the conjugate gradient technique.  In general, a preconditioner is an
approximation to $\left ( \rm{P^T N^{-1} P} \right )^{-1}$ that is used
to speed the convergence of the conjugate gradient solution.  In
general, we would use a two-piece preconditioner, with separate parts
for the sky solution and the correlated noise component.  However, if
the columns of $\rm{A}$ are diagonal (which they are by construction),
the natural preconditioner for the correlated part of the noise is
simply the identity matrix and so we neglect it.  The sky
preconditioner is a diagonal matrix with elements equal to the inverse
of the ``$n_{\rm obs}$'' map --- a map of the number of observations
in each pixel.  We find the use of this preconditioner greatly aids in
the convergence of the conjugate gradient solution for \artwo\ data,
unlike for the 148 GHz data where it makes little difference.

%
%
%
%
%
%
\begin{figure*}[!ht]
\plotone{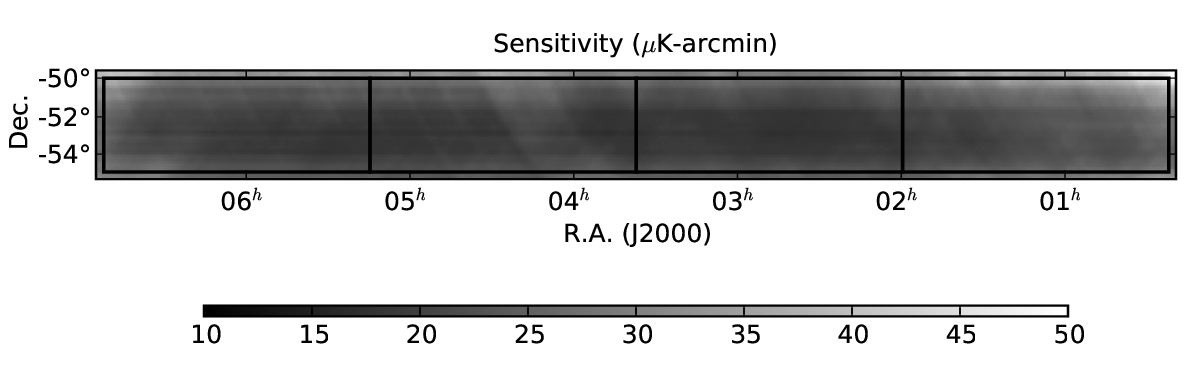}
\plotone{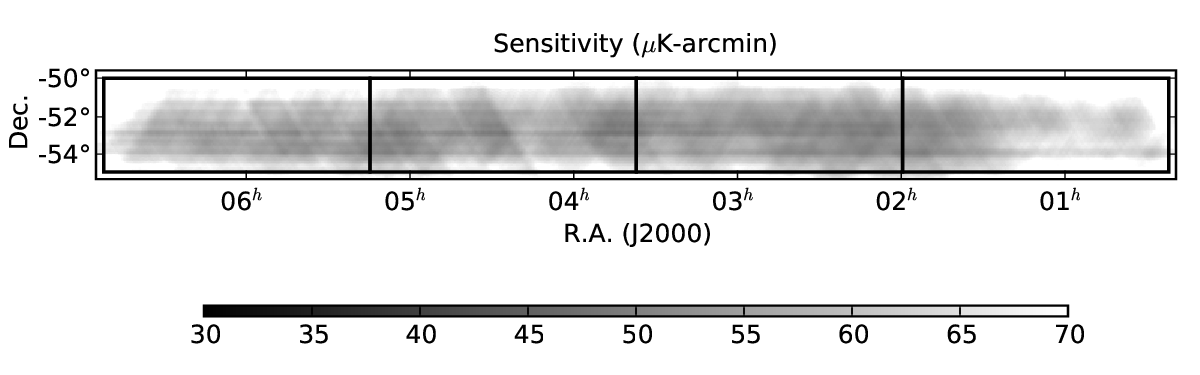}
\caption{The rms temperature uncertainty for one-arcminute pixels for the \arone{} maps (top) and \artwo{} map (bottom). 
Also shown in bold lines are the boundaries of the four rectangular patches used for spectral analysis.\label{fig: sensMaps}}
\end{figure*}
\subsection{Beam Measurements}
The window function due to the beam shape is obtained in a method similar to that followed in \citealt{Hincks/etal:prep}, hereafter H09. 
We use 22 (26) maps of Saturn from throughout the 2008 season to estimate the beam shapes for \arone\, (\artwo). The maps are made with an
independent pipeline from that used in \citetalias{Hincks/etal:prep}, but produce consistent results. Each map is binned into a symmetrized radial beam
profile, with 0\farcm17 bins, out to a radius, $\theta^{\rm b}_{\rm max},$ of 8\arcmin (6\arcmin) for \arone\,  (\artwo). 
 The dominant noise source in the maps comes from the slow variation of atmospheric brightness, and 
 since this projects into relatively large angular scales it causes significant noise correlation between radial bins in each profile.  The individual profiles
 are thus used to estimate the mean profile, and the full covariance matrix between bins is estimated from the scatter between them.  This is in 
 contrast to  \citetalias{Hincks/etal:prep}, where the mean profile was computed from a stacked map, and only diagonal errors were considered.  The full  covariance formulation serves to propagate the large scale map noise into low-$\ell$ uncertainty in the window function.
 The zero level of the radial profile is poorly constrained in the maps due to atmospheric
contamination at large radii. Because the sharp circular edge of the cold Lyot stop produces an Airy pattern at large scales, the radial profile is
expected to fall as $1/\theta^3$ \citepalias{Hincks/etal:prep}, so a fit to this form beyond $\theta^{\rm b}_{\rm max}$ estimates the zero level
and its covariance with the binned profile. The resulting function is also used to extrapolate the beam when computing the beam transform. 

Following the same prescription as \citetalias{Hincks/etal:prep}, a small set of basis functions is fitted to the inner part of the beam profile. A natural
choice of basis functions in harmonic space are Zernike polynomials which are compact on the unit disk, or their transforms in angular space which are Bessel functions divided by a linear function of radius.
For \arone\,  (\artwo) only 25 (18) basis functions are required to approximate the beam profile at the 1\%  (2\%) level within 1\arcmin\  and below the 15\% level within 8\arcmin ( 6\arcmin). Using a
small set of basis functions simplifies calculation of the covariance matrix of the window function  \citep{page/etal:2003} from the covariance of the
coefficients of the basis functions and the wing fit.
For the purpose of calibration against WMAP maps \citep{hajian/etal:prep}, the normalization of the window function is fixed to unity at $\ell =700$, and the calibration
error is factored out of the covariance matrix (described in H09) so that the window function has zero error at $\ell=700$.
Between $\ell = 1000$ and $10000$, uncertainties in the window function at \arone\ are between 0.7 and 0.4\%, and at \artwo\ between 1.5 and 0.7\%.  The uncertainty in the window function is significantly higher for $\ell < 700$, where atmospheric contamination makes measurements very difficult. In this paper, 
we use the beam window function only. The full covariance matrix of the window function is used in \citet{dunkley/etal:prep}.

%
%
\subsection{Calibration}
ACT scans every point in the survey area both when it is rising and setting. This cross-linking along with the unbiased 
map-making method described in Section~\ref{sec: mapmaking} allows the reconstruction of all modes in the maps, without biasing the 
large-angle modes. 
 As described in \citet{hajian/etal:prep},  we calibrate the \arone\ ACT maps directly to WMAP 7-year 94~GHz maps  \citep{jarosik/etal:prep}
 of the identical regions. Thus, the cosmological analysis is done with the same data used for the calibration.
 By matching the ACT-WMAP cross-spectrum  to the ACT power spectrum and the 
 WMAP 7-year power spectrum  \citep{larson/etal:prep} in the range $400<\ell <1000$, 
 \citet{hajian/etal:prep} calibrate the \arone\ ACT spectrum to  $ 2\%$ fractional temperature uncertainty.
 Similar methods are applied to \artwo\ maps, but their larger map noise levels on large angular scales mean that the 
 result does not currently improve on the results of an independent method of calibration, based 
 on observations of Uranus\footnote{It may be noted that although we use observations of Saturn for evaluating beam shape, we do not use them for calibration. Reliable modeling of the brightness temperature of Saturn is complicated by the effects of its rings.  The effective temperature changes with the ring inclination over the season. This is not the case for Uranus, whose brightness temperature is  fairly stable.}, described below.
 
 ACT made approximately thirty usable observation of the planet Uranus during its 2008 season.  We generated a map per observation after 
 calibrating the time-ordered data to detector power units, and determined the peak response of the planet with corrections for temperature
 dilution due to the finite instrumental beam size. The result is then compared to the Uranus temperature  in CMB differential units at the effective 
 band center to obtain a calibration factor \citep[see][for more details]{hajian/etal:prep}.  We take the brightness temperature of Uranus to be $107 \pm 6$ K and $96 \pm 6$ K for the
  \arone{} and \artwo{} bands respectively.  These temperatures 
 are based on a reprocessing of the data presented by \cite{griffin/orton:1993}, in combination with WMAP 7-year measurements of Mars and Uranus 
 brightnesses \citep{weiland/etal:inprep}. We find an uncertainty of  $7\%$ in the Uranus-based calibration that is dominated by the $6\%$ 
 uncertainty in the planet's temperature. The absolute calibration is consistent with the WMAP-based calibration
 described above. For \artwo{}, we adopt the Uranus-based calibration. 
 
%
%
\section{POWER SPECTRUM METHOD}
\label{sec: ps_method}
\citet{das/hajian/spergel:2009} describe the basis of our power spectrum method.  This paper improves on the analyses of 
\citet{fowler/etal:prep} by  optimizing
 the angular frequency resolution of the spectrum over the damping tail to resolve the acoustic features, as well as 
 by  extending  the multipole range of the power spectrum to $\ell=10000$. 
\begin{figure}[!htpb]
\begin{center}
\includegraphics[scale=0.4]{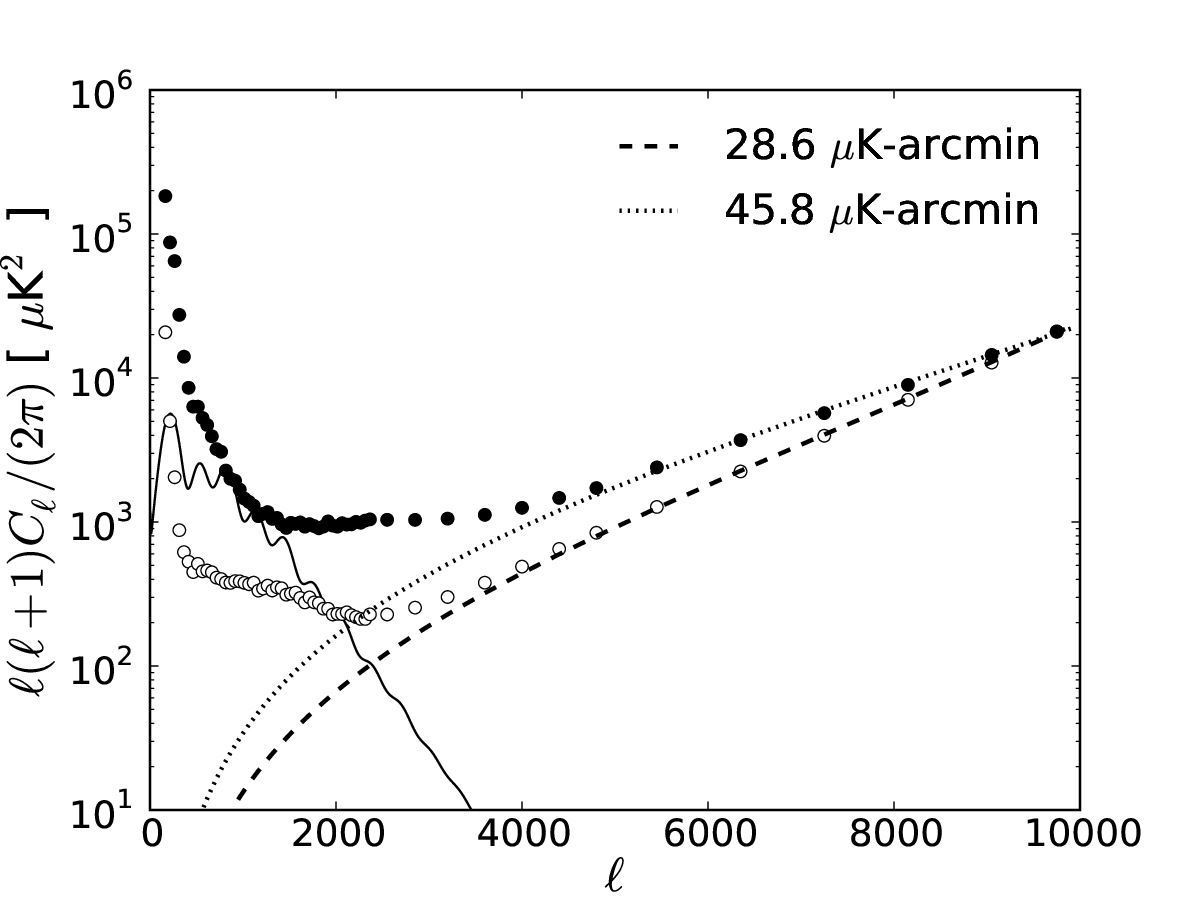}
\end{center}
\caption{Beam deconvolved noise power spectra of the \arone{} (open circles) and the \artwo{} (filled circles)  maps shown against a theoretical  lensed CMB spectrum (solid line). The theoretical white noise levels are shown using dashed (for \arone{}) and dotted (for \artwo{}) lines.
\label{fig: sigNoise}}
\end{figure}
\subsection{Fields used for Spectra}
\label{ssec: spectraFields}
  
 The power spectrum is computed from independent analysis of \npatchtext  contiguous patches in the ACT southern stripe (see 
Fig.~\ref{fig: sensMaps}) that together cover a rectangular area  of $296 \deg^2$ from $\alpha =$ $\hmin{00}{22}$  to  $\hmin{06}{52}$
(5\fdeg 5 to  103\degree) in right ascension and from $\delta = {-55}\degree$  to ${-50}\degree$ in declination. Each patch is 
$5\degree \times  14\fdeg8$ in size.
The full dataset is also split into \ndivtext subsets by
distributing data from each of \ndivtext successive nights into different subsets, thereby producing \ndivtext independent maps of
identical coverage and roughly equal depth.  We refer to each subset as a  \emph{split}.  Therefore, for each of the two  frequency bands,
 we have \npatchtext patches with \ndivtext  splits in each patch.  Each split is properly cross-linked, i.e., it represents the maximum 
 likelihood  solution of a dataset containing  both rising and setting observations.  A typical white noise level in the \arone\ season map (all 
 four splits taken together) range
 from 25--50 \micro\kelvin-arcmin, while for the \artwo\ map it varies between 40 and 70 \micro\kelvin-arcmin, with the noise being higher
 toward the edges of the maps (Fig.~\ref{fig: sensMaps}). The noise power spectra of the \arone\ and \artwo\ maps are shown in Fig.~\ref{fig: sigNoise}. 
 At small scales, the noise is typically white, while at larger scales there is a plateau of noise with a red tail towards low multipoles,
which is primarily attributable to atmospheric contamination.

\subsection{Preprocessing of Maps}
\label{ssec: preprocess}
Before patches are cut from the splits, each split map is high-pass filtered in Fourier-space. 
The high-pass filter is designed to remove all modes below $\ell=100$, and suppress the modes between $\ell=100$ and 500
with a sine-squared function, which rises from zero to unity within that range. The very lowest ($\ell<100$) multipoles are
dominated by extremely large atmospheric noise, and rolling off the filter to $\ell=500$  prevents these modes from leaking into and 
unnecessarily biasing the higher multipole portion of the spectrum. \par
Next, we \emph{prewhiten} the maps. Prewhitening   is  a local, real-space operation on the map
 designed to reduce the dynamic range of its Fourier components \citep{das/hajian/spergel:2009}.
 Prewhitening is particularly important for the current spectrum, as it is designed to specifically target the damping tail  of 
 the CMB ($1000<\ell<3000$).  
 This region of the spectrum has a steep slope ($C_\ell \sim \ell^{-4}$).  In the absence of prewhitening, finite boundary 
 effects and application of the point source masks cause a large amount of power to be aliased from low to high multipoles
 contaminating the spectrum at large multipoles. Although an unbiased estimate of the spectrum is obtained by deconvolving
  the mode-coupling window from the spectrum  (See Section~\ref{ssec:Mode-coupling} and \citealt{das/hajian/spergel:2009}.), this contamination
  adds to the uncertainty on the estimate, 
  resulting  in unnecessarily large error bars.  Prewhitening is performed by taking the difference of two versions of the same map, one convolved with a disc of radius 1\arcmin\ and the other with a disc of 3\arcmin.  To simulate discs with a pixelated kernel, we use the cloud-in-cell approach where we 
  assign appropriate weights for pixels under the disc according to its area of overlap with the disc.   This operation approximately amounts to taking 
  the Laplacian of the map, which 
  effectively multiplies the map Fourier transform by $\sim (\ell R)^2$  and the spectrum by $\sim (\ell  R)^4$ in the
  multipole range $900<\ell<3000$. Here $R = 1\arcmin$
   is the radius of the smaller disc.  This flattens out the damping tail but makes the  less colored lower multipole 
  portion of the spectrum steeply rising. This transition occurs around $\ell \simeq 900$, where $\ell^2 R^2 \approx 0.02$. 
  Therefore, we add  2\% of the original map back to the resulting map and this flattens out the $\ell<900$ portion of the spectrum.
  Note that the prewhitening kernel is designed  to flatten out by $\ell \sim 3500$ and therefore leaves the high multipole
  composite tail (which is already white)  largely unaffected. 
    
  We perform these actions in real space, but calculate the resulting prewhitening transfer function in Fourier-space so we 
  can undo the prewhitening in the final map estimate.  
 Prewhitening reduces the scatter in the $\ell=2000-4000$  region of the power spectrum, which provides significant statistical weight for extracting information on the SZ and correlated point source signals. 
\subsection{The Data Window}
\label{ssec: dataWindow}

After the maps are prewhitened, four patches are cut from each split. Each split patch is then multiplied with a window. The window
 is a product of three components ---  a  point source mask, the $n_{\rm obs}$ map, and a tapering function.  The details of the point source mask are given in Section~\ref{sec:fg}. To simplify the application, we generate a single  $n_{\rm obs}$ map per patch for each frequency
 by summing over the \ndivtext  splits for that patch. For the cross-frequency spectra we again generate a single  
  $n_{\rm obs}$ map by summing the $n_{\rm obs}$ maps from the two frequencies. 
 Multiplication by the $n_{\rm obs}$ map  essentially downweights the relatively scantily observed 
 and poorly cross-linked pixels toward the top and bottom edges of the map. Finally, to avoid ringing from the sharp edges of a patch, each 
 patch is multiplied by a taper that gently rolls to zero at the edge of the map.  We use a simple taper generated by taking a tophat 
 function which is unity in the center and zero over ten pixels at the edges, and convolving this with a Gaussian kernel of  FWHM $\simeq 2.5\arcmin$. \par
 
 We will denote the two frequency channels as $A$ and $B$, and use $i,j,k,l$ to denote the sub-season 
 splits. The patches will be denoted by Greek indices.  We will distinguish the  data windows only by their patch indices, and  denote them by
  $W_{\alpha}(\nhat)$, suppressing the channel labels. Thus, split $i$ of the patch $\alpha$ of frequency $A$,  after windowing, becomes\footnote
  {Throughout the paper, we use $X(\nhat)$ to denote a real space quantity, and $X(\bl)$  or $X_\bl$ to denote its Fourier transform.
  We use the tilde to denote windowing (not Fourier transformation), so that 
 $\widetilde X(\nhat)$  is $ W(\nhat) X(\nhat)$ after multiplication 
   by a window.  All calculation is done under the flat-sky approximation. }:
 \be
 \label{eq: windowedT}
 \widetilde T^{iA}_{\alpha}(\nhat) \equiv W_{\alpha}(\nhat) T^{iA}_{F;\alpha}(\nhat) 
 \ee
 where $T^{iA}_{F;\alpha}(\nhat)$ denotes the  map after filtering and prewhitening, 
 and is related to the original map $T^{iA}_{\alpha}(\nhat)$ through the Fourier-space relation:
 \be
 \label{eq: transferFunc}
 T^{iA}_{F;\alpha}(\bl)  =  F_\ell T^{iA}_{\alpha}(\bl), 
\ee 
where $F_\ell$ is a product of the prewhitening and beam transfer functions, the pixel window function and 
the high-pass sine-squared filter.  Here, and in what follows, the boldface $\bl$ is meant to represent a 2D wave-vector 
in Fourier-space, while we use regular $\ell$ to represent its the absolute value.  Symbols subscripted by $\ell$ 
represent quantities that are isotropic, while ones shown with an argument $(\bl)$ or subscripted by $\bl$, 
stand for explicitly 2D quantities in Fourier-space. 
  %
 %
 \subsection{Azimuthal Fourier-mode Weights and Binning \label{ssec: azWeights}}

\begin{figure}[!tpb]
\epsscale{1.1}
\plotone{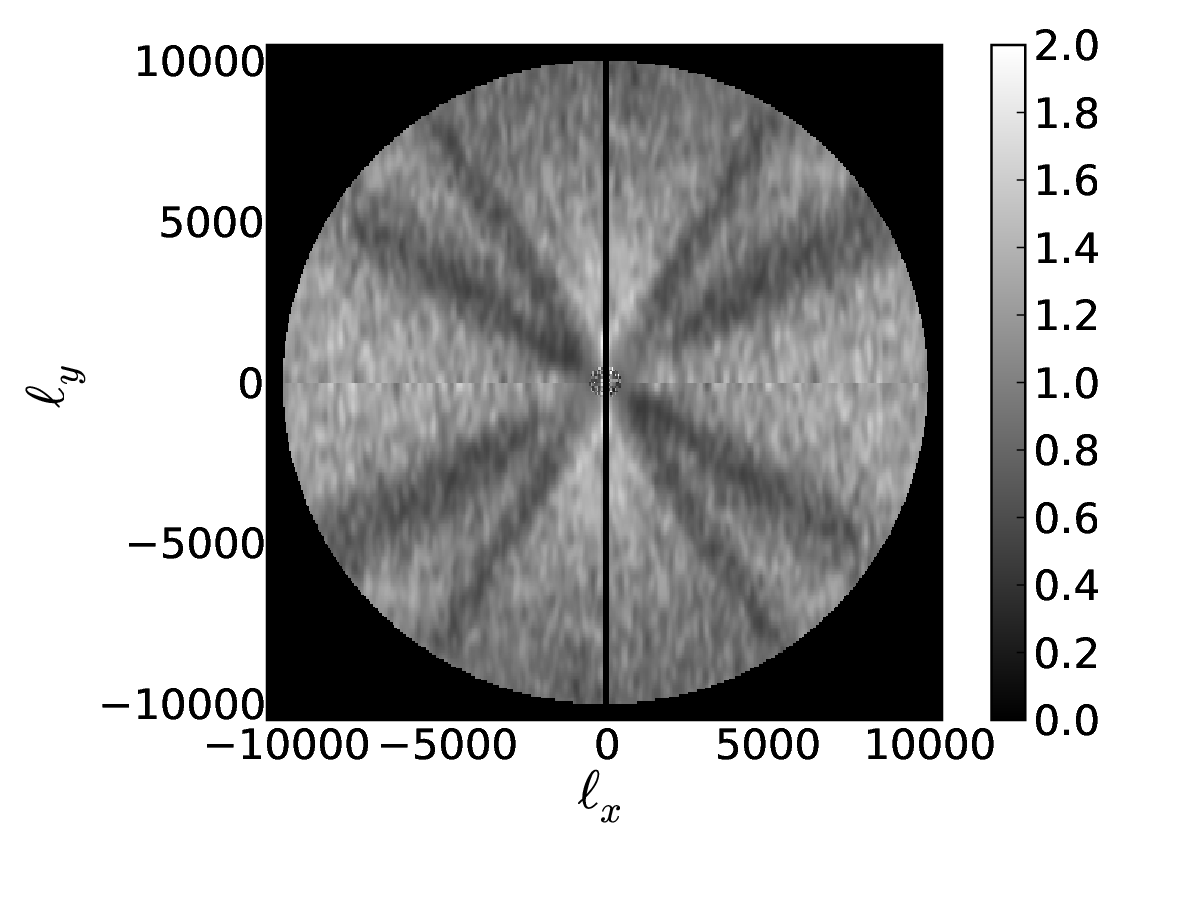}
\caption{Azimuthal weights used for binning the $148\times 148$ GHz spectrum. The horizontal stripe 
is an artifact of the symmetry of the Fourier space, and does not affect our calculations. 
The vertical stripe of zero weight is added in by hand, as described in the 
text.  Due to the symmetry of  Fourier transforms, only the upper half plane 
of this weight map is independent. The small 
grainy disk at the center corresponds to  $\ell<500$  where we do not perform any smoothing ---
 these modes are discarded 
from the final power spectrum.  \label{fig: azWeights}}
\end{figure}

After windowing the data, a 2D pseudo-spectrum is computed as:
\be
\label{eq: pseudoCl}
\widetilde C_\bl^{iA\times jB} = \mathrm{Re}\left[\widetilde T^{*iA}_{\bl} \widetilde T^{jB}_{\bl} \right], 
\ee
where the patch index has been suppressed to simplify notation. 
At this stage, we crop out and retain a rectangular area of  the 2D spectrum defined by 
$-\lend<\ell_x<\lend$ and $-\lend<\ell_y<\lend$, throwing out all information
at $\ell_x,\ell_y > \lend$.   This trimming  reduces the number $n_F$
 of Fourier-space pixels by a factor of  $\sim 4$, and is crucial for several subsequent operations, 
 especially the direct computation of the mode-coupling matrix (see Section~\ref{ssec:Mode-coupling}),
 which involves a computational step that scales as $n_F^2$. 
 The 1D binned spectrum $\widetilde C_b$ results from averaging the 2D spectrum in 
 annular bins:
 \be
\label{eq: binnedPseudoCl}
\widetilde C_b^{iA\times jB} = \sum_{\bl} P_{b\bl} \widetilde C_\bl^{iA\times jB},
\ee
where $P_{b\bl}$ is the binning matrix, whose value is zero when $\bl$ is outside the annulus
defined by bin index $b$.  Because the noise properties are not isotropic in Fourier 
space, we employ a 2D weighted average, with the weights determined by the noise power spectrum. 
We briefly describe here how the weights are computed. For a given frequency, say, $A$, we compute the mean 
2D auto spectrum from the four split patches and subtract the mean 2D cross-power spectrum from it. 
This gives an estimate of the 2D noise power spectrum for the split patch $N_\bl^{AA}$. We assume 
that each split has roughly equal noise, and approximate the season noise power spectrum as
$N_\bl^{AA}/\ndiv$. We estimate the 2D variance of the spectral estimator $C_\bl^{iA\times jB}$ as
\be
\label{eq: sigmaSqCl}
 \sigma^2(C_\bl^{iA\times jB}) \simeq (C_\bl^{\rm th} {b^A_\ell}^2 +N_\bl^{AA}/\ndiv)(C_\bl^{\rm th} {b^B_\ell}^2 +N_\bl^{BB}/\ndiv) ,
 \ee
 where $C_\bl^{\rm th}$ is  an azimuthally symmetric theoretical CMB spectrum, and $b_\ell$ represents the beam transfer function. 
 The first estimate of the azimuthal weights
  is chosen to be the inverse of the variance (Eq.~[\ref{eq: sigmaSqCl}]):
 \be
 w^{iA\times jB}_\bl = \frac 1 { \sigma^2(C_\bl^{iA\times jB}) } . 
 \ee
The resulting 2D \emph{Fourier-space map}\footnote{The 2D Fourier transform $X(\bl)$  of a real-space map $X(\nhat)$
 is defined on a 2D pixelated grid in $\bl$-space.   We  use the term ``Fourier-space map'' to denote any 
 2D quantity defined on the same $\bl$-space  grid.  Such a quantity can be solely defined in $\bl$-space and 
 is not necessarily derived from a Fourier transform of some known real-space quantity.} of weights
  is typically noisy, and before using it to bin the power spectrum, we treat it in the following manner.
 In every annulus, defined by the bins $b$, we identify outlier pixels whose values are greater than a threshold $\alpha_b$
 (typically chosen to be 10 times the median in that annulus) and set their values to  $\alpha_b$. 
 We then divide the pixels in each annulus by their mean, and smooth the resulting 2D Fourier-space map with a 
 Gaussian with FWHM of three pixels. These operations are designed to produce well-behaved  Fourier-space weight maps,
  with emphasis on bringing  out the overall azimuthal variation in the weights by removing outliers and radial dependence. 
 The Fourier-space weight map used in the \arone$\times$\arone{}  spectrum is shown 
 in Fig.~\ref{fig: azWeights}.  A comparison of this Fourier-space weight map with the signal and noise levels in Fig.~\ref{fig: sigNoise} shows
 the expected behavior of the weights being isotropic in the signal-dominated regime at low $\ell$, with the anisotropy becoming more pronounced
 as the noise begins to dominate. The main sources of anisotropy in the weights are the rays of excess noise along and
 perpendicular to the scan directions, leading to two slightly rotated (because the rising and setting scans are not exactly orthogonal)
 ``X'' shaped patterns of low weight regions.  Another significant noise term is attributable to scan synchronous noise, likely caused by instabilities
 induced by acceleration at scan turn-arounds. The noise leads to horizontal striping in the maps and manifests itself  as an excess of power  
 in a vertical strip $-90 < \ell  <90$ in the Fourier space. We set  the  pixels  inside this strip in the 2D  Fourier-space weight
  map to zero. In terms of the 2D weights, the binning matrix can be expressed as, 
 \be
 \label{eq: binningMatrix}
 P_{b\bl} = \left. \frac{w_\bl}{\sum_\bl{w_\bl}} \right|_{\bl \in b} .
 \ee
  It is noteworthy that  both the  $n_{\rm obs}$ weighting in real space and the azimuthal weighting in Fourier space
   lead to  a 5-10\% percent lowering of the uncertainties in the final spectral estimates. 
%
%
\subsection{Mode-coupling \label{ssec:Mode-coupling}}
From Equations (\ref{eq: windowedT}) to (\ref{eq: binnedPseudoCl}), we can express the binned 
pseudo-spectrum  $\widetilde C_b$ in terms of the underlying spectrum $C_\bl$:
\be
\label{eq: binnedPseudoVsTheory}
\widetilde C_b^{iA\times jB} = \sum_{\bl,\bl' } P_{b\bl}  |W(\bl -\bl')|^2 F_{\ell'}^2 C^{iA\times jB}_{\bl'}.
\ee
 This quantity can be related to a binned version of the the true spectrum $C_b$ through an  inverse binning 
 operator $Q_{\bl b}$,
 which is unity when $\bl \in b$ and zero otherwise:
 
 \ba
\nonumber \widetilde C_b^{iA\times jB} & =& \sum_{\bl,\bl',b' } P_{b\bl} |W(\bl -\bl')|^2  F_{\ell'}^2 Q_{\bl' b'}C^{iA\times jB}_{b'}\\
& \equiv & \sum_{b'} M_{bb'} C^{iA\times jB}_{b'},
\ea
 where $M_{bb'}$ is the mode-coupling matrix.
 We compute the $M_{bb'}$ exactly without resorting to any of the one-dimensional approximations
  often used in the flat-sky case \citep{hivon/etal:2002, das/hajian/spergel:2009,lueker/etal:2010}. 
  The mode-coupling matrices are well behaved and stable to inversion. 
We define the unbiased, decorrelated estimator of the power spectrum (indicated with the circumflex):
\be
\label{eq: specEstimator}
\widehat C_b^{iA\times jB} =\sum_{b'}  M^{-1}_{bb'}  \widetilde C_{b'}^{iA\times jB}.
\ee
 \begin{figure*}[!ht]
 \epsscale{1.1}
\includegraphics[scale=0.45]{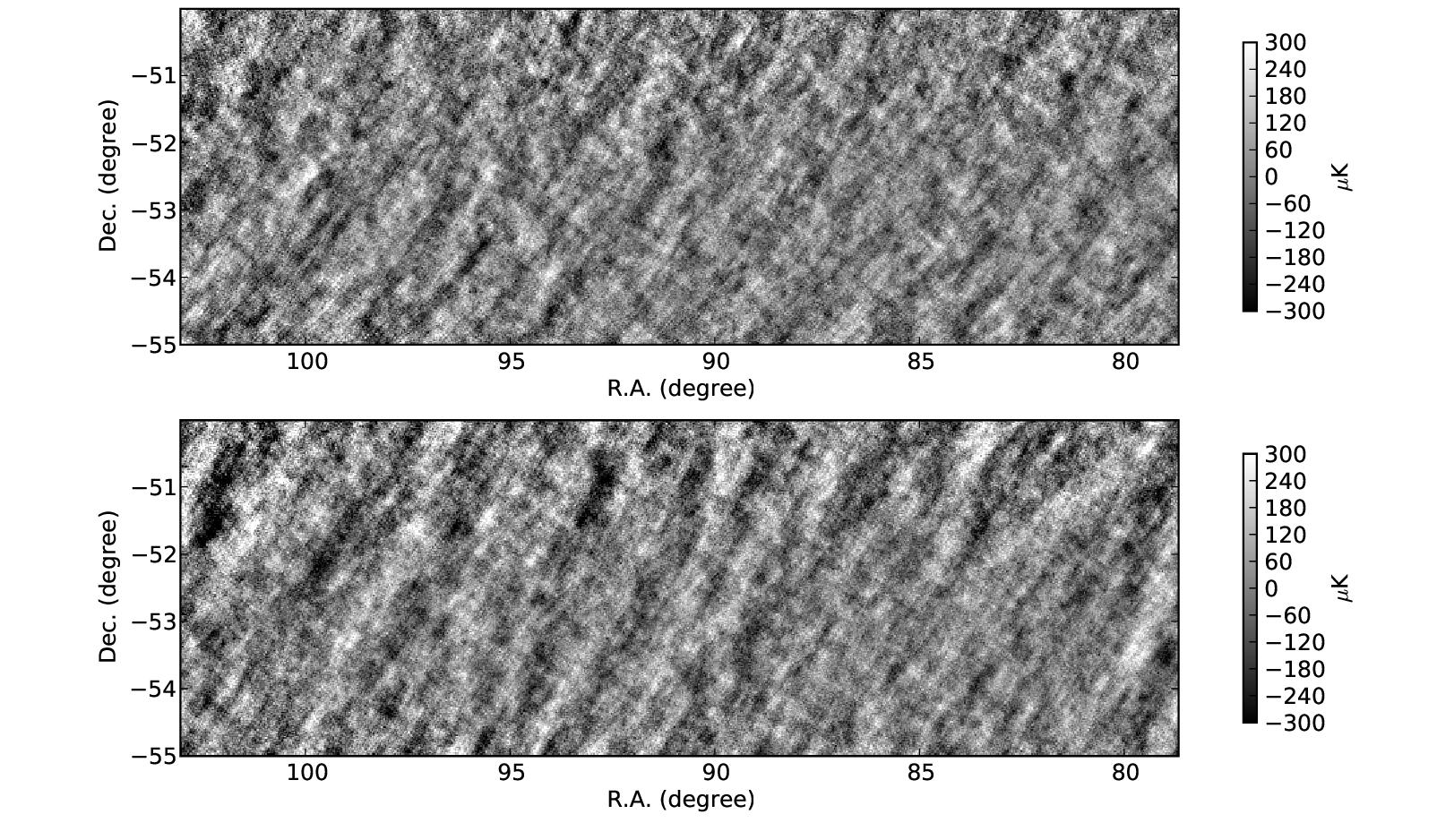}\includegraphics[scale=0.45]{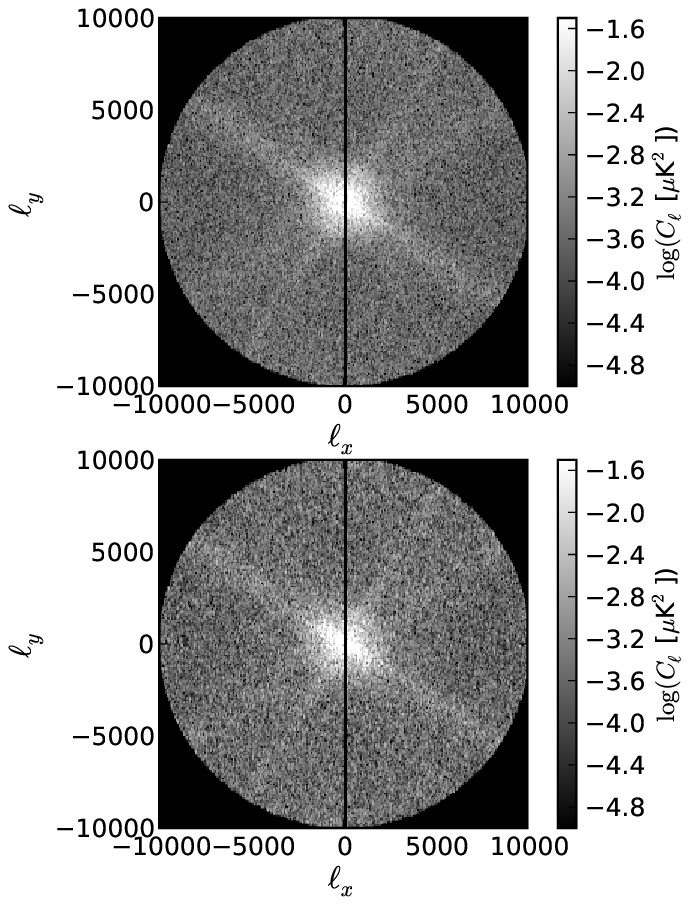}
\caption{Anisotropic noise in data and simulated patches. \emph{Top}:  Difference map of two splits in an \arone\ data patch (\emph{left}), and
 its 2D power spectrum (\emph{right}).  
\emph{Bottom}: Same as above, except for a simulated patch. This random realization was seeded by the noise power spectrum 
of data patch shown in the top panel, as described in the text.
  \label{fig: randomNoiseComparison}}
\end{figure*}
For a single frequency spectrum,  we compute the cross-power spectrum in each patch as the mean of the six cross-power spectra
 (from four splits).  We also compute the auto-spectrum as  the mean of four auto-spectra from the splits. 
 The variance for a given patch is determined by combining the noise estimate (obtained as one-fourth
  the difference of the mean auto and the
 mean cross-spectrum) in that patch with a theoretical
 signal  power spectrum.
  We combine the four mean cross-power spectra from the  patches with  inverse variance weighting to obtain the final spectrum.
  \par
 For the cross-frequency power spectra, we compute twelve cross-power spectra in each patch by crossing each split for a frequency 
 with the three other splits from the other frequency. We do not cross the same splits (consisting of 
 data from the same nights) between the two frequencies 
 because of the possible contamination from covariant atmospheric noise among the two channels. 
 We proceed by averaging those twelve per patch, and then combining the cross-spectra from the patches 
 with inverse weighting as above. In this case, we estimate the variance in a patch as the square root of the 
 product of the single-frequency variances in that patch. 
 \par
  For parameter estimation \citep[see][]{dunkley/etal:prep}, we wish to compare our spectral estimates $\widehat{C}_b$ to theoretical power spectra
   $C_\ell^{\rm th}$. To do so we find the bandpower  window function, $B_{b\ell}$ which converts  $C_\ell^{\rm th}$ to binned theoretical spectra
    $C_b^{\rm th}$.  (We suppress the superscripts on the $C$s for clarity.)  We begin with Equation~(\ref{eq: binnedPseudoVsTheory}) which relates the
    binned spectrum to the underlying 2D  spectrum, and introduce a matrix $I_{\bl \ell}$
    that describes the mapping of a theoretical spectrum defined at integer multipoles onto our discrete Fourier pixels.  
The binned  theoretical spectrum follows from Equation~(\ref{eq: specEstimator});
\ba
\nonumber  C_b^{\rm th} &=&  \sum_{b',\bl'',\bl'}  M^{-1}_{bb'} P_{b'\bl''}  |W(\bl'' -\bl')|^2 F_{\ell'}^2 I_{\bl' \ell} ~C_\ell^{\rm th}
 \\& \equiv & B_{b\ell} ~C_\ell^{\rm th}.
\ea
Since each rectangular Fourier-space pixel spans a range of integral multipoles,  it gets contributions from several multipoles in the theoretical
spectrum. We choose $I_{\bl \ell}$ as a  Gaussian with its peak at the pixel center $(\bl_x,\bl_y)$,  and  width corresponding to the pixel size, so as
to give more 
 weight toward the center of the pixel.  The bandpower window functions are insensitive to the exact form of this response function.  As with the mode-coupling matrix, we  compute $B_{b\ell}$ exactly.

\subsection{Summary of power spectrum method}

The angular power spectrum is conceptually simple, but measuring it over
a wide range of angular scales with sufficient angular resolution to see
acoustic peaks from maps over a small portion of sky
with significant variation in sensitivity over the map, is technically challenging.
As a quick reference, we summarize here our  multiple-step procedure to accomplish this task:
\begin{itemize}
\item  Divide of the data into
independent ``splits'' each comprising a quarter of the total observation nights,
and construct independent maps for each split;
\item High-pass filter  the maps, to eliminate the large angular scale modes which
are poorly constrained and can bleed power into the smaller-scale modes;
\item Prewhiten the maps through real-space convolutions so that the
power spectrum of the maps is flattened, which reduces aliasing of
power from large scales to small scales given the steep power spectrum of the maps;
\item Divide each split map into four independent patches and construct
a window for each patch which accounts for varying sensitivity over the patch,
perform masking of point sources, and edge tapering
to avoid spurious power from a sharp cutoff; 
\item  Calculate 2D
binned pseudo-spectra from the Fourier transformation of the windowed, prewhitened,
filtered maps, for each pair of frequency, split, and patch values; 
\item Construct  a one-dimensional binned pseudo-spectrum via
azmiuthal averaging of the two-dimensional spectrum over bins in multipole $\ell$; 
\item  Estimate the true binned power spectrum from the binned 
one-dimensional pseudo-spectrum by inverting a mode-coupling matrix
which accounts for the effects of beam profile, prewhitening, filtering, 
pixelization, and windowing. 
\end{itemize}
We combine the spectra from individual patches with inverse 
variance weighting to obtain the final spectrum. 
We also construct a bandpower window function
which converts a theoretical power spectrum into the corresponding binned
power spectrum which we estimate from the maps. We verify these techniques
on sky simulations, described below.

\section{SIMULATIONS}
\label{sec: simulations}
Simulations play an important role both in verifying that our mapper is working correctly and in understanding our spectrum. 
 Specifically, we use simulations to test for map convergence, estimate bandpower covariances, and confirm analytic 
estimates of the power spectrum uncertainties.\par
To test for convergence, we perform a set of end-to-end simulations of the \arone\ and \artwo\ maps.
We simulate signal-only time streams by mock observations of input celestial maps (from \citealt{sehgal/etal:2010})
 and generate noise time-streams which capture the main features of the noise in the data.  
\begin{figure}[!b]
\epsscale{1.2}
\plotone{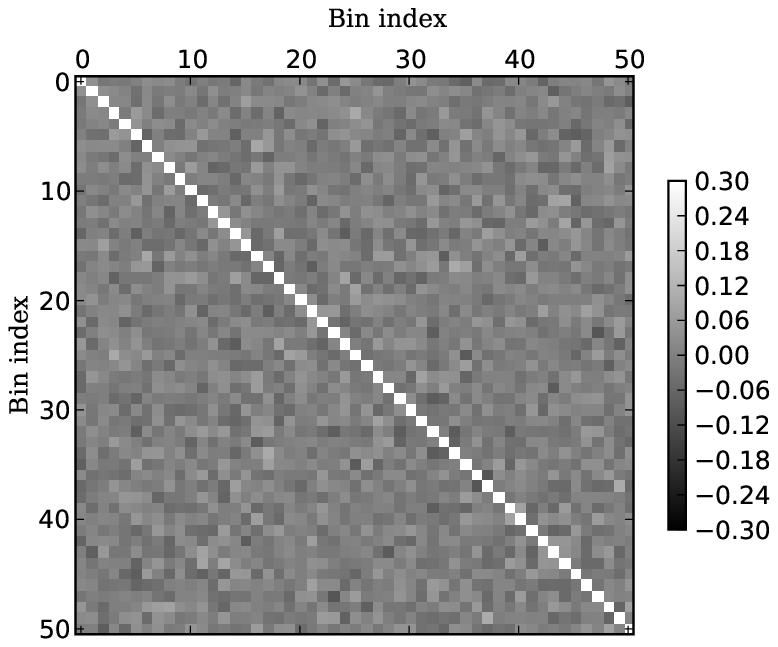}
\caption{Normalized covariance matrix for decorrelated 
bandpowers for \arone\ $\times$ \arone\ spectrum, 
  based on 960 signal+noise Monte Carlo simulations. The bins are defined 
in Table~\ref{tab: spec_table}.  All values on the 
diagonal are unity by definition. The colorbar has been stretched to reveal the 
variations in the off-diagonal elements. \label{fig: 148_covariance}}
\end{figure}
 To test for convergence, we simulate a signal-plus-noise map, subtract from it a noise-only map at the same 
 iteration  and compare the power spectrum of the  resulting difference map with that from a signal-only map.
 For \arone, the maps are well converged to multipoles as low as $\ell \gtrsim 200$ by iteration 1000, and we keep $\ell \ge 500$
 in our final spectrum.  For the \artwo\ maps, where higher atmospheric
noise warranted a slightly different mapmaking approach (see Section~\ref{sec: mapmaking}), convergence 
at large scales is slow.  After 1000 iterations, the \artwo\ maps are converged
down to $\ell=1000$, but not at smaller $\ell$.    Moreover, the \artwo\ spectrum is highly noise-dominated at low multipoles.
  Thus, we limit spectra  involving \artwo\  to $\ell\ge 1500 $ in  our analyses.  

\begin{figure*}
\plotone{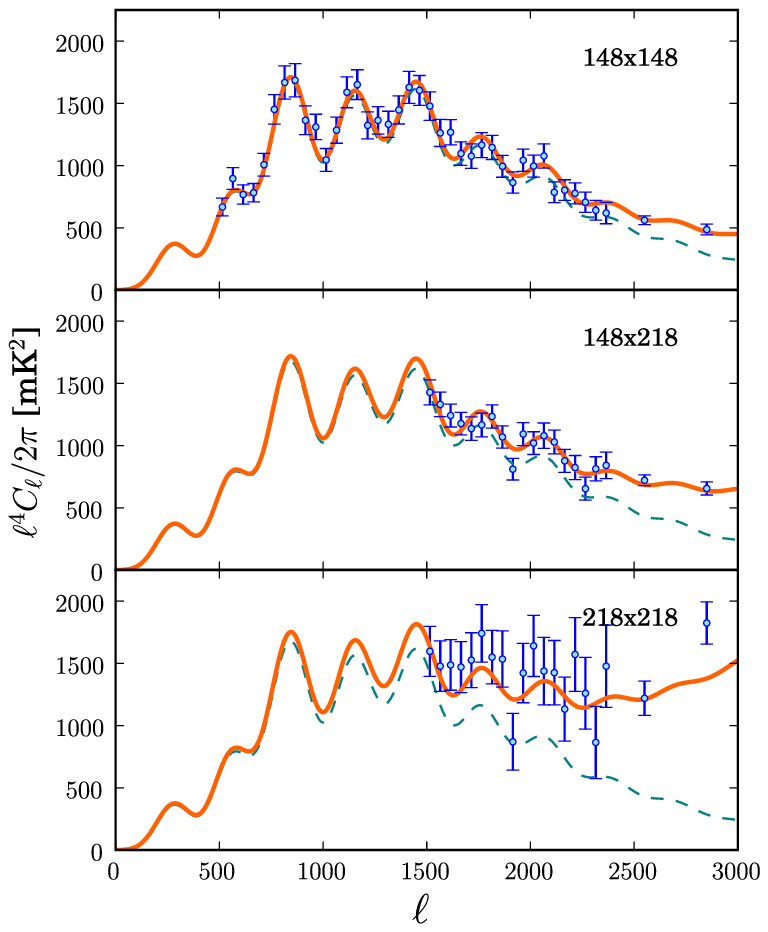}
\caption{Intermediate multipole $500<\ell<3000$ portion of the power spectra plotted with an $\ell^4$ scaling to emphasize the acoustic peaks.
 The thick orange 
curve shows the best-fit model including the CMB secondaries and point source  contribution
 taken from our companion paper \citet{dunkley/etal:prep}. The model depends on the
 frequency combination. 
 The thin dashed teal line shows the best-fit lensed CMB-only theory, which is frequency
 independent.  
 From top to bottom, the three panels show the \arone, the \ara$\times$\artwo\ and  the \artwo\ spectra.
 For this plot, data spectra and errors  from Table~\ref{tab: spec_table} have been scaled by best-fit calibration factors from \citet{dunkley/etal:prep}
  of $1.02^2$ , $1.02 \times 1.09$ and $1.09^2$ for the \arone,  the \ara$\times$\artwo\ and  the \artwo\ spectra, respectively.
 \label{fig: ell4} }
\end{figure*}
\par
 In order to investigate the bandpower covariance and to validate the analytic prescription for 
uncertainties in the spectrum, we ran a large set of Monte Carlo simulations that contain key
aspects of the noise properties manifested in the data.  Due to the iterative nature of our map-making 
pipeline, it is prohibitively expensive to run a large number of end-to-end simulations starting from 
simulated timestreams. Therefore, we adopt an intermediate  approach, in which we find a prescription 
from the data for generating realistic realizations of noise in map space, to which we add signal realizations.
We start with two quarter-season split maps, $T_0(\nhat)$ and $T_1(\nhat)$, with corresponding
$n_{\rm obs}$ maps labeled $n_0(\nhat)$ and $n_1(\nhat)$. We use the difference of these maps 
to estimate a one-hit noise map  $D_{01}(\nhat)$ 
(i.e.,  a noise map representative of the variance in each pixel had it been observed only once), 
\ba
 D_{01}(\nhat) = \frac{T_0(\nhat)-T_1(\nhat)}{\sqrt{{n_{0}^{-1}(\nhat)}+{n_{1}^{-1}(\nhat)}}}.
\ea
With the other two splits, we compute another one-hit noise map $D_{23}(\nhat)$.
We compute the average of the 2D power spectra of  $D_{01}$ and  $D_{23}$, 
and use its  square root  as the  
 amplitude for a Gaussian random field, and generate random one-hit noise maps by randomizing the phases
and converting back to real space. 
Finally, we divide one such map by $\sqrt{n_i(\nhat)}$ to generate a realization of the noise for split $i$. 
This procedure captures the non-white and anisotropic aspects of  map noise, as shown by 
Fig.~\ref{fig: randomNoiseComparison}. 
\par
\begin{figure*}
\plotone{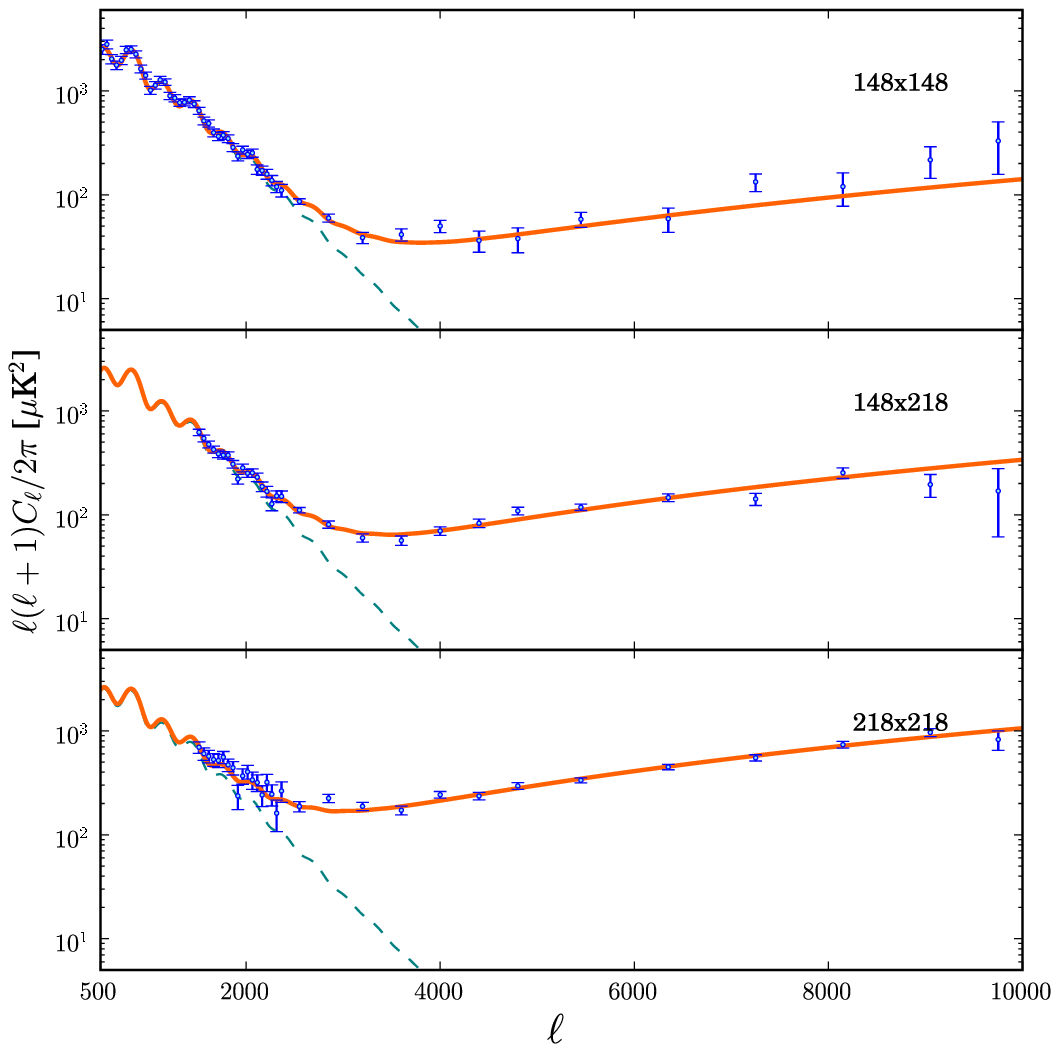}
\caption{Single and cross-frequency spectra plotted with the conventional $\ell(\ell+1)$ scaling,
 highlighting the behavior at large multipoles. The thick orange 
curve shows the best-fit model including the CMB secondaries and point source  contribution
 taken from the companion paper \citet{dunkley/etal:prep}.  The model depends on the
 frequency combination. 
 The thin dashed teal line shows the best-fit lensed CMB-only theory, which is frequency
 independent.     From top to bottom, the three panels show the \arone, the \ara$\times$\artwo\ 
  and  the \artwo\ spectra.
  For this plot, data spectra and errors  from Table~\ref{tab: spec_table} have been scaled by best-fit calibration factors from \citet{dunkley/etal:prep}
  of $1.02^2$ , $1.02 \times 1.09$ and $1.09^2$ for the \arone,  the \ara$\times$\artwo\ and  the \artwo\ spectra, respectively.
  \label{fig: ell2}}
\end{figure*}
We also simulate a random residual point source component in our Monte Carlo realizations. 
The point source model for \arone\ consists of two components - a radio source population
and a dusty sub-mm population, both based on \citet{toffolatti/etal:1998}.  Source maps are generated
assuming a Poisson distribution.  For each flux bin $S$, spanning the range $0.01-20$ mJy, we generate 
the number of point sources in that bin as a Poisson realization from the model flux distribution $dN/dS$
(the upper limit is chosen to be the approximate $5\sigma$ detection threshold at \arone). Once the number
of sources in each bin is generated, their individual fluxes are dithered to redistribute the fluxes within each bin, and the sources
are laid down at random positions in the map.  For \artwo\ we scale the radio and IR components by their 
appropriate spectral indices, based on the findings of \citet{dunkley/etal:prep}. With these settings, we 
closely reproduce the  level of  the Poisson point source signal 
seen in the high multipole regime of the data power spectra. We estimate the Poisson contribution 
to the error bars on bandpowers from the scatter of 500 simulations of point source-only maps, and add that 
as a correction to the (signal+noise)-only error-budget obtained from the signal. At the current level of
noise in our maps, these corrections are largely negligible, and we ignore their contribution in  the analytic 
error bar estimation  and band power covariance. \par
\begin{figure*}[!ht]
\plotone{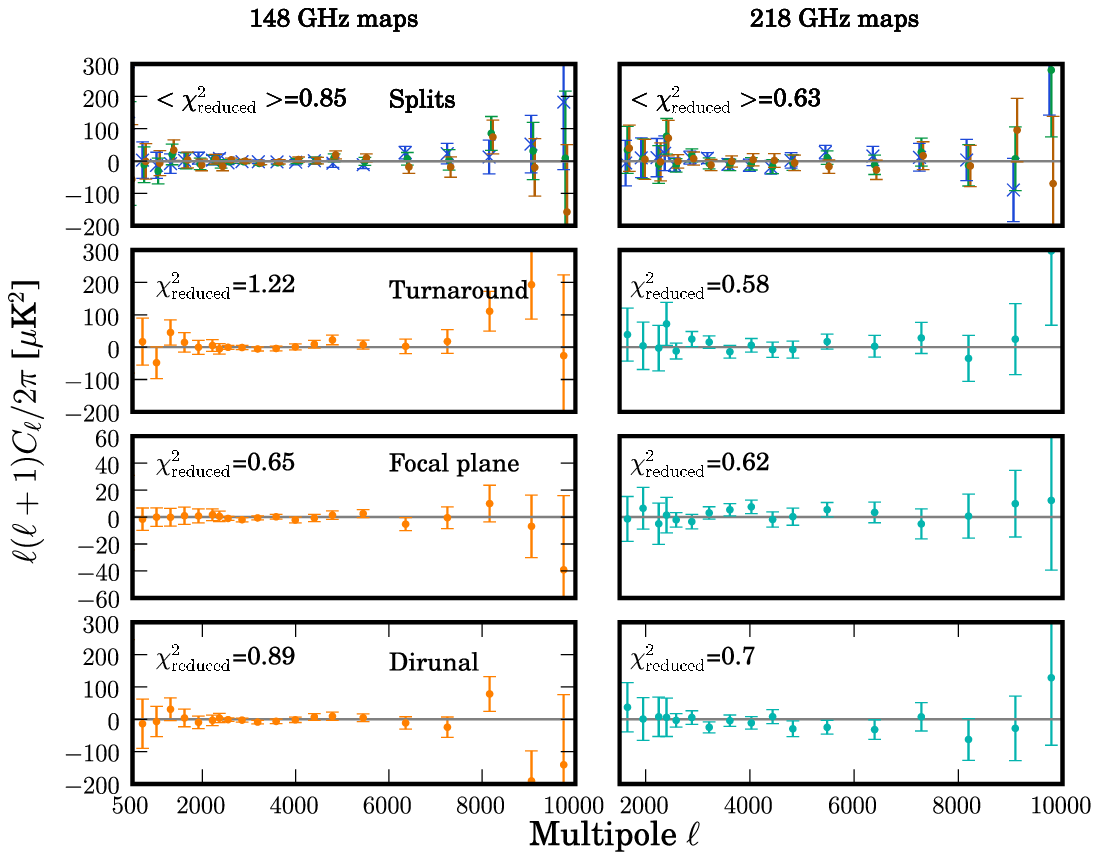}
\caption{Null tests of the ACT 148 (left) and 218 (right) GHz maps. The top row plots the null spectra from difference maps made from the four 
different time splits described in Section~\ref{ssec: spectraFields}. The second row illustrates the systematic check on whether data near the 
telescope turnarounds are  contaminated, while the bottom two rows check for a systematic gain difference across halves of the focal plane and for diurnal effects, 
respectively.  The difference spectra and the chi-squares are computed with the same bins as the standard power spectra.
 For display purposes only, we have re-binned the spectra in this plot with larger bins. 
 The spectra are consistent with null in all cases.\label{nullspeck}}
\end{figure*}
We estimate \arone\ and \artwo\ spectra for 960 signal+noise realizations. Each realization 
involves simulating four patches and four splits in each patch for each frequency, and estimating  
the spectra with exactly the same methods as used on the data. 
The resulting spectra are used for quantifying the covariance between band powers.
Fig.~\ref{fig: 148_covariance} displays the normalized covariance matrix,
 ${\rm Cov}(i,j)/\sqrt{{\rm Cov}(i,i) {\rm Cov}(j,j)}$, where $i, j $ denote 
 bin indices,  showing
that correlations between adjacent bins are insignificant at the 10\% level.

The uncertainties in the band powers are evaluated using the analytic prescription described in the Appendix.
 First, a large set of isotropic white noise simulations are generated to validate the analytic formula for the single frequency and the cross-frequency error bars.
These simulated error bars are found to be in agreement with analytic prescription to less than a percent. 
The next set of simulations are run with anisotropic noise, Poisson point source realizations, and with real-space and Fourier-space weighting 
as described in Section~\ref{sec: ps_method}, and small corrections to analytic prescription due
 to the anisotropic nature of the noise and weighting are evaluated against the Monte Carlo simulations.
 
\section{POWER SPECTRUM }
\label{sec: results}
 Applying the methods described in Section~\ref{sec: ps_method}, we compute 
 the decorrelated bandpowers for the \ara, \arb\ and \ara$\times$\artwo{} 
 spectra from the unbiased map solutions. The bandpowers are presented 
 in Table~\ref{tab: spec_table} and are displayed in Figures~\ref{fig: ell4} and ~\ref{fig: ell2}.
 In Fig.~\ref{fig: ell4} we have plotted the lower multipole portion of the power spectrum
 $\lbegin<\ell<3000$ with an $\ell^4$ scaling to emphasize the higher-order acoustic 
 peaks in the primary CMB spectrum. In the \arone\ spectrum
 we clearly resolve the second through the seventh peaks of the CMB. As discussed 
 in \citet{dunkley/etal:prep}, the higher-order peaks provide  leverage on cosmological 
 parameters such as the spectral index and its running, the primordial helium fraction and  the number
 of relativistic species. Fig.~\ref{fig: ell2}, on the other hand,  emphasizes the high multipole 
 tails of the spectra, where the signal is dominated by emission from dusty star-forming 
 galaxies and unresolved radio sources.  The intermediate range $2000<\ell<4000$ gets a significant 
 contribution from the thermal  Sunyaev-Zeldovich effect, and lets us constrain 
 the amplitude of  the SZ spectrum. \par
  \begin{figure*}[!ht]
 \plotone{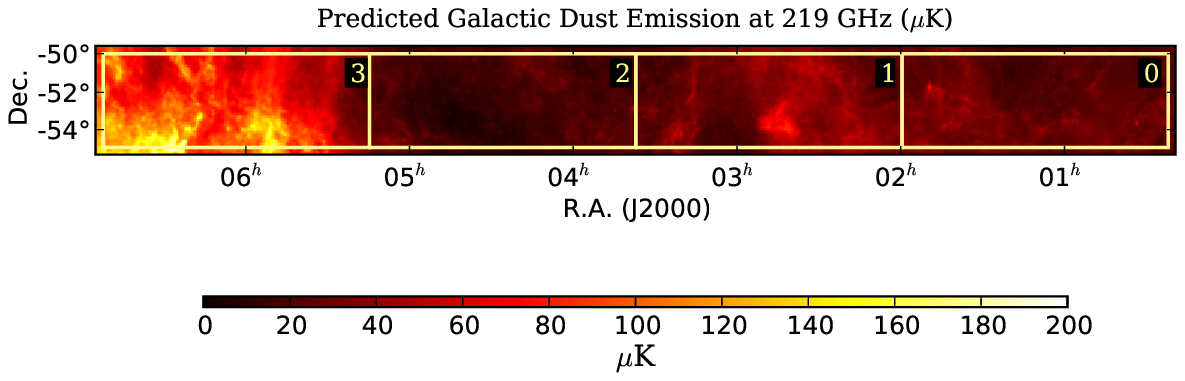}
 \caption{Galactic thermal dust emission template at 219 GHz based on \citet{finkbeiner/davis/schlegel:1999}``model 8".  We cross correlate 
 the template patches (numbered here) with ACT patches to look for a correlated dust signal in ACT maps.  Patches 0-3 are located 
 off the Galactic plane with central Galactic latitudes of $-64$\degree, $-56$\degree, $-43$\degree, and $-28$\degree, 
 respectively. For the dust emission the effective central frequency in the ACT \artwo\ band is 220 GHz \citep{swetz/etal:prep}. \label{fig: fdsMap}}
 \end{figure*}
The data are checked for consistency by performing various null tests, in which difference maps are
 constructed  to cancel true sky signals, and their power spectra examined.  We describe the suite of tests here.
As described in Section~\ref{ssec: spectraFields}, the TODs are split into four subsets, generating four maps with independent detector noise. We 
verify that these maps are consistent with each other by generating difference maps from each pair, and computing the two-way cross spectra from the 
three independent pairs of difference maps:
\begin{eqnarray}
C_\bl^{(i-j)\times (k-l)} = \ave{\Delta T^{(i-j)*} (\bl)  \Delta T ^{(k-l)} (\bl)}
\end{eqnarray}
where $\Delta T^{(i-j)} \equiv T^i -T^j$ and the indices $i,j,k,l$ range from 0-3.
The same point source mask  used to mask the full dataset is
applied to the difference maps before calculating the spectra; all other
settings also remain the same.
The difference maps, $\Delta T^{(i-j)}$, \
are downweighted by the same  $n_{obs}$  maps summed over the four splits  
used in the windowing of the CMB data (see Section~\ref{ssec: dataWindow}).
Similarly, when binning the power spectra, we use the same azimuthal 
weights in each patch described in Section~\ref{ssec: dataWindow}.
The three spectra are shown in the top two panels of Fig.~\ref{nullspeck} for the 148 (top left)
 and 218 (top right) GHz maps. We compute the bandpowers in the range $500<\ell<10000$ ($1500<\ell<10000$)
 for 148\,(218)~GHz. The
mean of the three spectra is consistent with null
with $\chi^2=42\,(14)$ for 148\,(218)~GHz,  with $51\,(31)$~degrees of freedom.

Another null test probes the consistency of data taken with the telescope accelerating as it reverses direction at the ends of the scan  (turnarounds). Note that for the standard maps, no turnaround cut is applied. 
Four new split maps are made cutting  data near the turnarounds, amounting to losing 
$\simeq 24\%$  of the total data. Two difference maps are made by pairing one split of the  standard map with a different split of the new maps (we
avoid differencing the same splits as they have very similar noise structure), and a two-way cross-power spectrum is produced.  Any artifact due to 
the turnaround would be left in these difference maps and might produce excess power. However, we find
the resulting spectra, shown in the second row of Fig.~\ref{nullspeck},  to be 
consistent with null with a $\chi^2 = 62$ and ${18}$ respectively for the 148, 218 maps (again for 51 and 31 degrees of freedom).

The last two null tests probe possible systematics associated with the focal plane and from diurnal effects.  The first compares maps made from only the top half of the detector array with those from the bottom half.  The second compares maps processed with data only from the middle half of the night to the standard maps.  The resulting spectra are shown in the final two panels of Fig.~\ref{nullspeck}, and are also consistent with null.

\section{FOREGROUNDS}
 \label{sec:fg}
The main foregrounds in the \arone{} and \artwo{} bands at the angular scales considered here are point sources. We
mask the detected ones as described below. Another 
foreground is the diffuse Galactic dust, which is discussed below in some detail. In the companion paper, 
\citet{dunkley/etal:prep}, we also consider the 
contributions to the power spectra from the thermal and kinetic SZ effects and the clustering of infrared point sources. \par
The point sources include radio sources and dusty star-forming galaxies (SFG).  The radio 
 sources are typically active galactic nuclei exhibiting  synchrotron-dominated spectra, with 
 emission extending down to lower radio frequencies. The dusty SFG are characterized 
 by absorption of ultraviolet photons from star-forming regions by dust.  The dust reemits into  graybody radiation
 that rises with increasing frequency into  sub-millimeter bands.\par
\begin{figure}
 \includegraphics[scale=0.45]{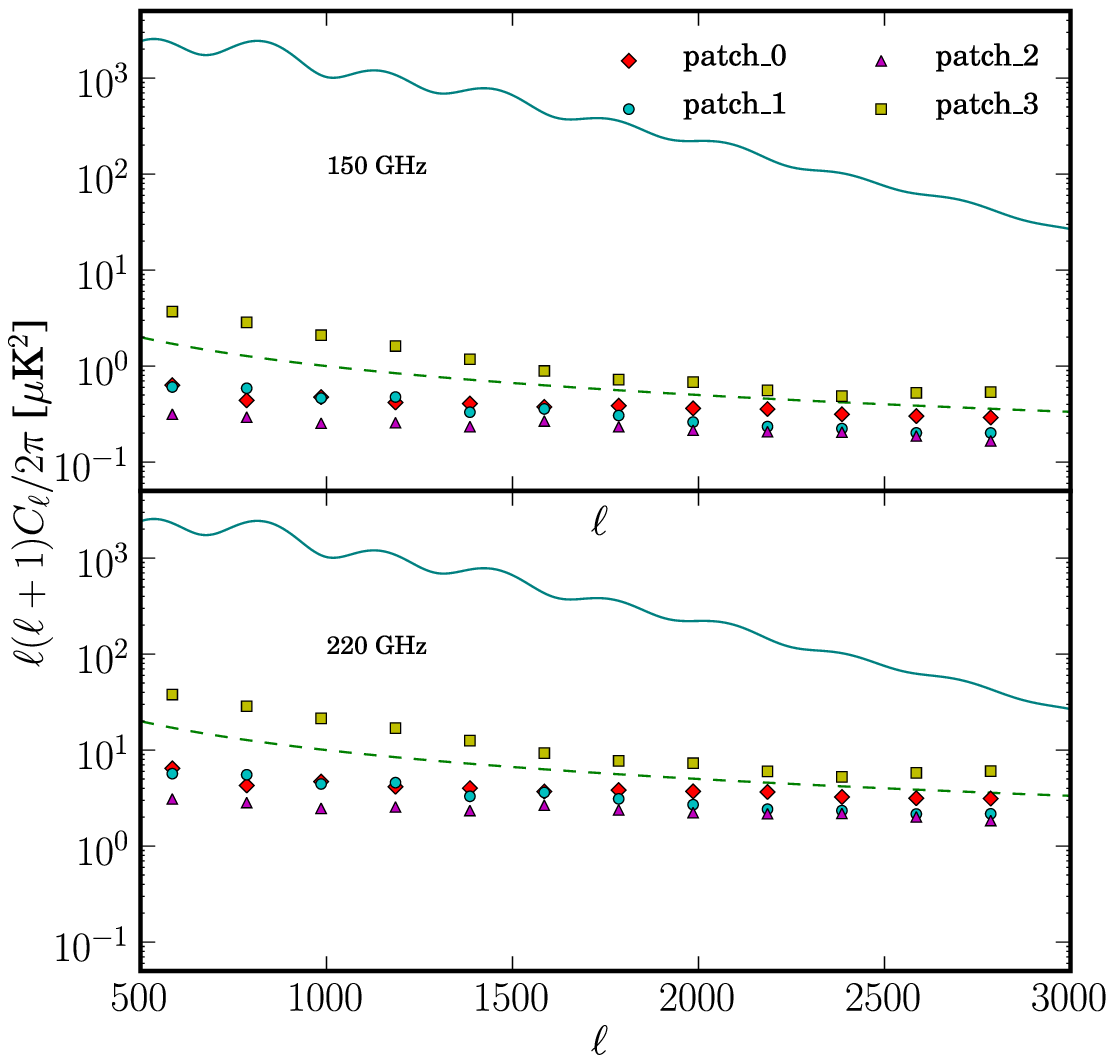}
 \caption{Power spectra of dust in the FDS template in the patches shown in Fig.~\ref{fig: fdsMap}  at \aroneDust\ and \artwoDust\ in CMB differential temperature units. 
 A theoretical lensed CMB power spectrum is shown for comparison  (continuous curve).
 Also shown is a $C_{\ell} \propto \ell^{-3}$ curve (dashed) normalized to $\ell^2 C_{\ell}/2\pi$ = 2 (20) \micro\kelvin$^2$
 at $\ell=500$ for \aroneDust\ (\artwoDust). Note that a temperature power law index of $\beta  = 1.7$ between \aroneDust\ and \artwoDust\
 corresponds to a factor of 10 in the power spectra. \label{fig: dustSpectra} }
 \end{figure}
For the purpose of extracting the underlying CMB signal, it is important to identify and mask bright point sources 
in the maps before the power spectrum is computed.   A detailed study of the point source population 
 detected in the ACT \arone{} 2008 survey is presented in \citetalias{marriage/etal:prepa}.
\begin{figure*}[!ht]
\plottwo{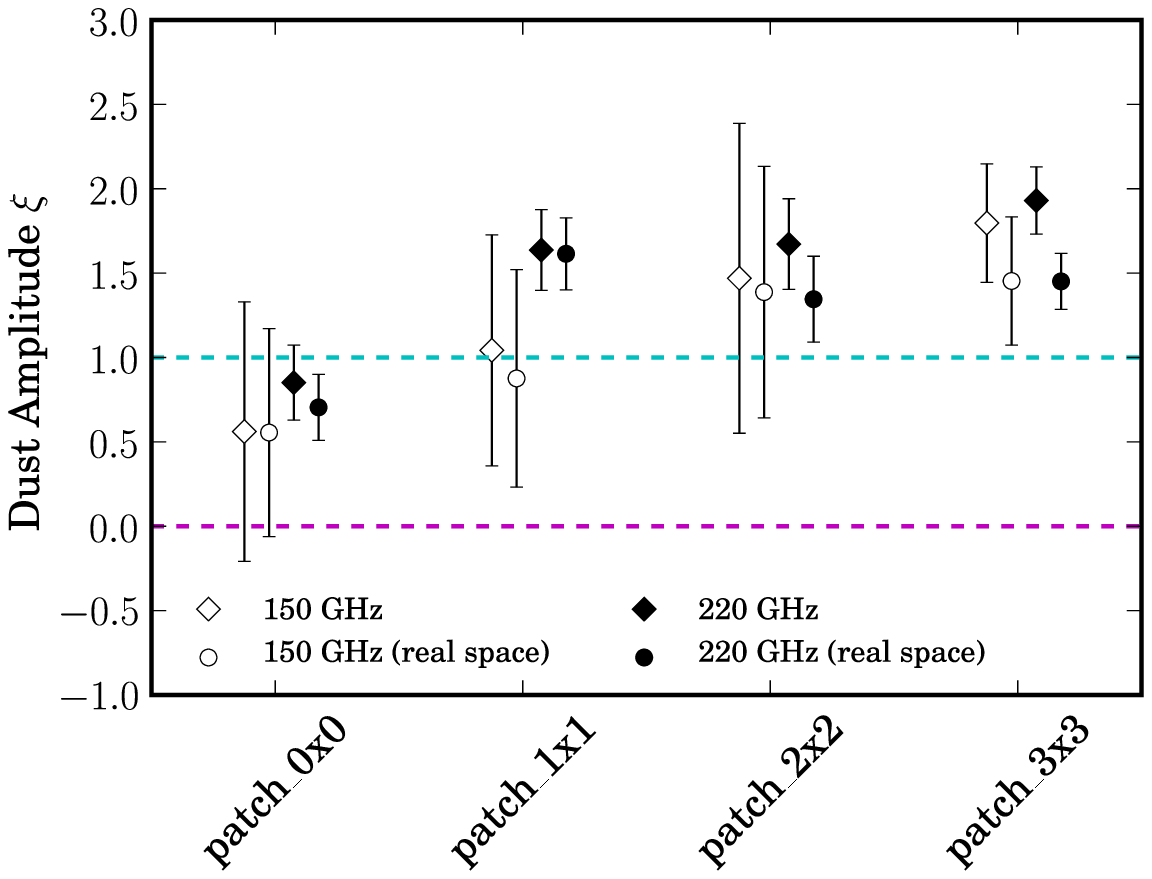}{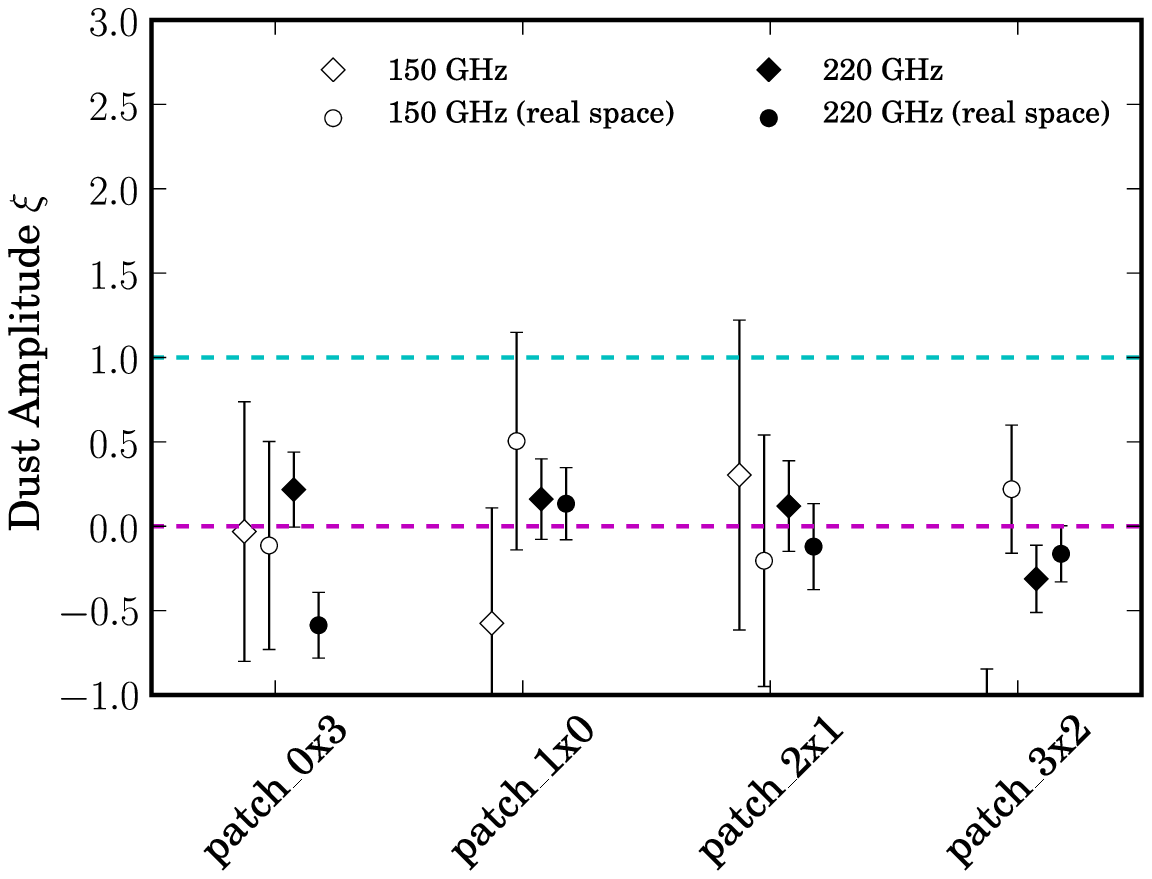}
\caption{{\emph{Left}}: Amplitude, $\xi$, of dust emission in ACT patches relative to FDS predictions  for \aroneDust\ (open symbols) and \artwoDust\ (closed) symbols. 
Results obtained through the real-space and Fourier-space techniques are shown with circles and diamonds, respectively. {\emph{Right}}: 
Results of a null test performed by shuffling around the FDS patches, so that patch 0 of ACT is crossed with patch 3 of FDS etc.
\label{fig: dustAmplitudes}}
\end{figure*}
Point sources 
are detected with the matched filter method \citep{tegmark/deOliveira-Costa:1998,wright/etal:2009,vieira/etal:2010}.
To construct the point source masks, we  consider 
 detections with SNR $>5 $ in either frequency.   Over the $296 \deg^2$ area used for power spectrum
estimation,  
we find a total of 164 point sources, taking the union of the detections at the two frequencies.  
Before computing  the spectrum, we mask out a  10\arcmin{} diameter region around each of the 164 point sources.
This amounts to masking about $1\%$ of the total area. As discussed in  Section~\ref{ssec: preprocess},
the prewhitening operation effectively guards against the leakage of power due to the application of this mask. 
\par

We estimate the Galactic dust contribution to our power spectra by
cross-correlating ACT maps with the predictions for infrared cirrus emission at CMB  frequencies 
based on the multicomponent dust model ``8'' of  \citet{finkbeiner/davis/schlegel:1999} (hereafter FDS).
\defcitealias{finkbeiner/davis/schlegel:1999}{FDS} The FDS-based dust maps at \arone\ and 219\,GHz are available
as a  part of the \citet{sehgal/etal:2010}  simulations\footnote{These simulations are available at 
\url{http://lambda.gsfc.nasa.gov/toolbox/tb_cmbsim_ov.cfm}
To be precise, we use the actual frequencies for which the FDS maps were made, and the effective
dust frequency for ACT, but in practice, the differences are negligible.
}.  We resample these maps into ACT pixelization, and cut out the ACT patch  regions.
The 219\,GHz FDS map is shown in Fig. ~\ref{fig: fdsMap}.  Patches 0 through 3 labeled in the 
figure are situated off  the Galactic plane with central Galactic latitudes of $-64$\degree, $-56$\degree,
 $-43$\degree and $-28$\degree, respectively. The power spectra of the predicted dust signal in each patch are shown
in Fig.~\ref{fig: dustSpectra}. 

We proceed with the cross-correlation as follows. We express the ACT map as a sum of the cosmic and galactic dust components:
$T_{\rm{ACT}} = T_{\rm{CMB}}+\xi ~T_{\rm{FDS}}$, where $\xi$ is the dust amplitude, 
predicted to be unity by FDS.  We take two approaches to estimate $\xi$: one based on real-space operations, 
and the other  based on power spectra. \par
In the real space approach, 
we prewhiten both the ACT and FDS maps, convolve the ACT maps with an $6\farcm1$ FWHM Gaussian beam appropriate 
to the FDS resolution, apply the ACT point source mask described above, filter out modes below $\ell =500$ (1500) for 
\aroneDust\ (\artwoDust) and produce the maps ${\cal T}^{i}_{\rm{ACT}}(\nhat)$ and ${\cal T}^{i}_{\rm{FDS}}(\nhat)$ in
patch $i$.   Then we estimate $\xi$ in patch $i$ as 
\be
\xi^{i}_{\rm real} =\frac{  \int d^2 \nhat ~ {\cal T}^{i}_{\rm{ACT}}(\nhat)  
{\cal T}^{i}_{\rm{FDS}} (\nhat) } {\int d^2 \nhat ~ {\cal T}^{i}_{\rm{FDS}}(\nhat) {\cal T}^{i}_{\rm{FDS}}(\nhat)}.
\ee
The error on the above estimate is obtained by cross-correlating 960 random CMB plus noise simulations 
(see Section~\ref{sec: simulations}) with the dust maps.  
In the spectrum-based approach, we estimate $\xi$ in each multipole bin  $b$ as the ratio of the ACT-FDS cross-spectrum to the auto-spectrum 
of the FDS map for the same patch:
\be
\xi_b^i = \frac{C_b^{{\rm ACT}\times {\rm FDS};i}}{C_b^{{\rm FDS} \times {\rm FDS};i}}  . 
\ee
The ACT-FDS cross-spectrum is calculated using both pixel-space weighting and  Fourier-space azimuthal weighting, as with the 
ACT power spectrum, with the FDS and ACT beams deconvolved.
The final estimate, $\xi_{\rm spec} ^{i}$ is obtained as the mean across the bins over the 
range $500<b <3000$ ($1500<b <3000$) for 
\aroneDust\ (\artwoDust) where  the upper 
limit is dictated by the  FDS resolution.  The 
uncertainty on $\xi_{\rm spec}^i$ is estimated as the scatter in the same quantity computed from the cross-spectra  between 
the FDS maps and  960 random CMB plus noise realizations. As a check for systematics, we circularly shift the FDS patches by one patch, 
such that patch\_0 of ACT is crossed with patch\_3 of FDS, patch\_1 of ACT with patch\_0 of FDS etc., and recompute the
real and Fourier space estimates of $\xi$.   The results are shown in Fig.~\ref{fig: dustAmplitudes}.
 The errors shown arise from treating the FDS template as deterministic. \par
 This figure shows that a dust-correlated signal is observed. More work will be needed to 
understand how well the FDS template represents faint dust well off the Galactic plane \citep[see, e.g.,][]{veneziani/etal:2010}.
The fact that such a small signal  ($\ell^2 C_\ell/2\pi \sim 0.1-1.0 ~\mu$K$^2$ at $\ell=3000$ and \arone) 
can be recovered with our map-based technique demonstrates the power of our maximum-likelihood map estimation. 
Recently  \citet{hall/etal:2010} have also reported a measurement of infrared cirrus in the SPT maps.
\par
Subtracting the FDS template from our maps has negligible effect on 
 the power spectra, consistent with the minuteness of the dust signal.

\section{LENSING OF THE CMB}
\label{sec: cmbLensing}
The CMB photons are deflected by large-scale structure potentials along their path from the
last scattering surface at $z\simeq 1100$ to us \citep[see][for a review]{lewis/challinor:2006}. The typical (rms) deflection
 in the $\Lambda$CDM model is about $2\farcm7$ and
the deflection pattern is coherent over degree scales, comparable to the size of the acoustic features on the primary 
CMB. These coherent deflections produce distortions of the hot and cold spots on the CMB, leading to a broadening 
of their size distribution. In the power spectrum, this effect manifests itself as the smoothing of the acoustic peaks, 
which can be used as a signal to look for lensing. The first attempt at a detection 
of lensing in the power spectrum 
was made in \citet{reichardt/etal:2009} using the data from the ACBAR experiment. They quantified the effect of including 
lensing in their analysis through the log-ratio of the lensed to unlensed Bayesian evidence ($\Delta {\ln \cal E}$), and by 
combining WMAP5 and ACBAR datasets found $\Delta {\ln \cal E}=2.63$. \citet{calabrese/etal:2008} analyzed the 
ACBAR data with a different approach, where they introduced a
scaling of the power spectrum of the lensing potential ($C_\ell^{\Psi} \rightarrow A_L C_\ell^{\Psi}$, with $A_L=0$ 
corresponding to no lensing and $A_L=1$ to the standard $\Lambda$CDM expectation). With WMAP5+ACBAR they  found
$A_L= 3.0\pmerrors{0.9}{0.9} $ (68\% confidence level). According to footnote 17 of  \citet{reichardt/etal:2009}, 
a reanalysis of the ACBAR data using this parameterization yields $A_L = 1.60\pmerrors{0.55}{0.26}$.
\par
\begin{figure}[!h]
\includegraphics[scale=0.45]{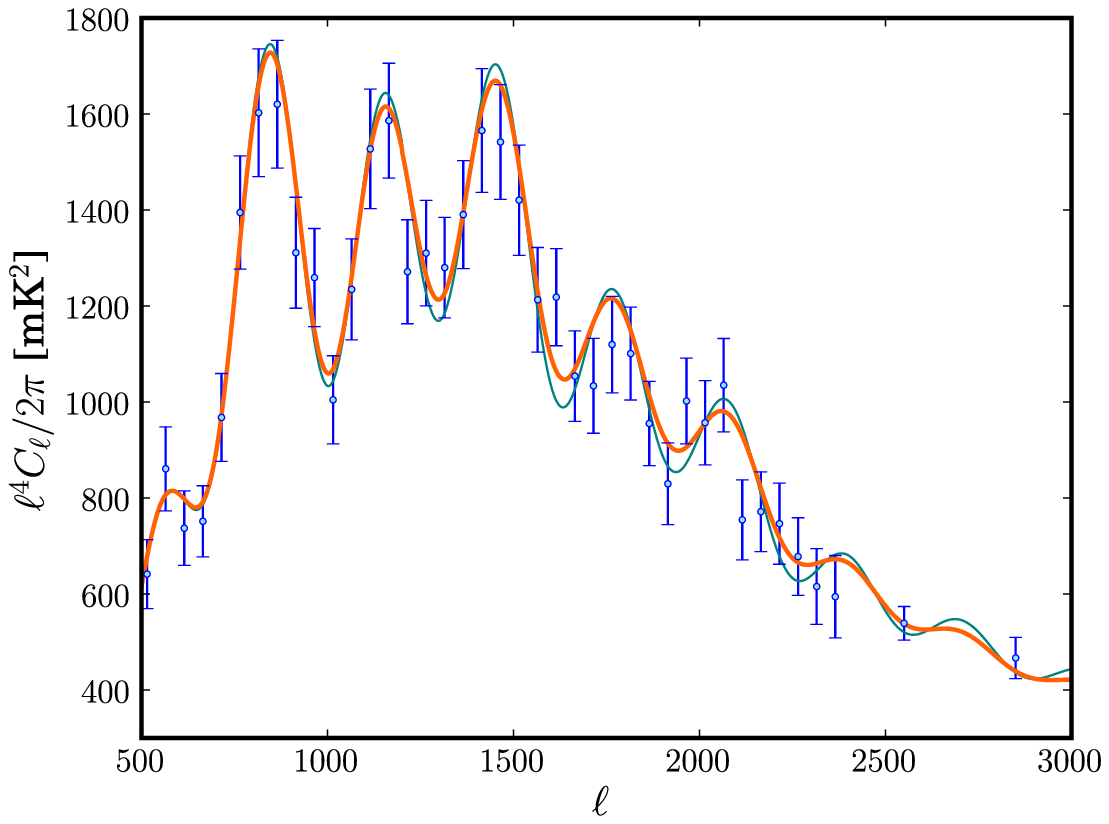}
\caption{Lensing of the CMB smooths out the acoustic peaks in the CMB power spectrum. The best fit model with 
 lensed CMB, secondaries, and point sources is
shown as the thick orange curve, while the same with no lensing is shown with the thin green curve.  \label{fig:lensedCMB}}
\end{figure}
\begin{figure}
\epsscale{1.2}
 \plotone{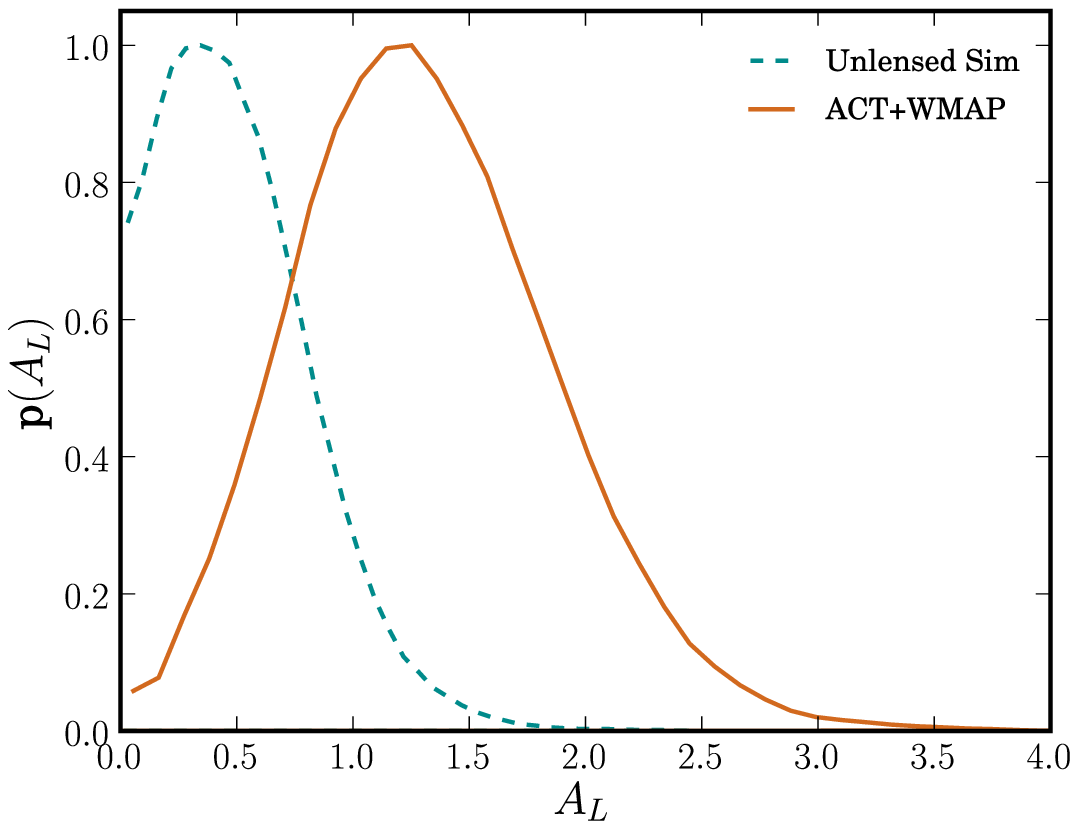}
 \caption{One dimensional marginalized distribution (solid line) for the lensing parameter $A_L$, that scales the expected lensing potential such that $C_\phi\rightarrow A_L~C_\phi$ \citep{calabrese/etal:2008}. An unlensed CMB spectrum would have $A_L=0$, and the standard lensing case has $A_L=1$. With ACT+WMAP7 we find $A_L=1.3^{+0.5+1.2}_{-0.5-1.0}$ (68\% and 95\% limits), a 2.8$\sigma$ detection of lensing. The dashed
 line shows the marginalized distribution for $A_L$, obtained from an unlensed  simulation. \label{fig: lensingConstraint}}
 \end{figure}

 
Our \arone\ power spectrum is shown against lensed and unlensed models in Fig.~\ref{fig:lensedCMB}.  
 We use the parametrization of \citet{calabrese/etal:2008}, and using the parameter estimation 
 methodology described in the companion paper \citet{dunkley/etal:prep}, we  constrain the lensing parameter
$A_L$ based on WMAP7 and ACT  power spectra. 
Fig.~\ref{fig: lensingConstraint} shows the marginalized 1D likelihood for $A_L$ using WMAP7+ACT. 
We find $A_L = 1.3^{+0.5 (+1.2)}_{-0.5 (-1.0)}$ at $68\% \,(95\%)$ confidence, with the best-fit lensed CMB spectrum
with $A_L=1$ 
 having an improved goodness-of-fit to the WMAP7+ACT data of $\Delta \chi^2 = 8$ less than the unlensed model.
\par
We check for systematics that might have given rise to a spurious lensing signal. The projection scheme (cylindrical-equal-area)
used for the ACT maps is not particularly optimized for lensing studies --- so we test whether this projection 
could  introduce a lensing-like signal.  We simulate a low-noise unlensed CMB signal and run it through the mapmaking pipeline, 
and try to reconstruct a ``lensing convergence'' in the resulting map using standard quadratic estimator techniques \citep{hu/okamoto:2002}.
 We find the reconstructed convergence power spectrum to be consistent with null (there is a small known bias  at high multipoles
   that we have entirely traced to mode-coupling due to the finiteness of the patch), showing that the projection does not 
 introduce any significant lensing-like signal in our patches. To further test if any other step in our pipeline could produce spurious 
 peak smearing in the power spectrum, we generate an end-to-end simulation of a noisy map exactly as described in Section~\ref{sec: simulations},
 only this time replacing the lensed CMB signal time-stream with its unlensed version.  The resulting maps are then processed through 
 the power spectrum pipeline, and the power spectrum is then analyzed with the parameter estimation method described in
  \citet{dunkley/etal:prep}.  We find the lensing amplitude parameter, $A_L$, described above to be consistent with zero (see Fig.~\ref{fig: lensingConstraint}).

Here we have used the smearing of the acoustic features to look for the lensing signal. More promising
ways of extracting this signal involve using information beyond the two-point correlation (optimal quadratic estimators)
and cross-correlation with tracers of large-scale structure.  Early efforts  cross-correlating the lensing reconstruction
in  WMAP 3-year maps with luminous red galaxies, radio sources and quasars
 \citep{smith/zahn/dore:2007, hirata/etal:2008} yielded $\sim 3\sigma$
evidence for lensing.  Data from WMAP are not the best suited for this kind of study because of its limited angular
resolution.  With arcminute resolution CMB data, such as from ACT, there is  a much higher potential for 
a detection. Such efforts are underway with the ACT maps.

\begin{deluxetable*}{cc|cc|cc|cc}
\tablewidth{0pt} 
\tablecolumns{7} 
 \tablecaption{Single frequency bandpowers\\ ${\cal B}_b = \ell_b (\ell_b+1) C_b/2\pi $   (\micro\kelvin$^2$) \label{tab: spec_table} }
\tablehead{\multicolumn{2}{c}{} & 
	\multicolumn{2}{c}{\arone} &
	\multicolumn{2}{c}{\arone{} $\times$ \artwo} &
	\multicolumn{2}{c}{\artwo}\\
	\colhead{$\ell$ range} & 
	\colhead{central $\ell_b$} &
	\colhead{${\cal B}_b$ }& 
	\colhead{$\sigma({\cal B}_b)$ }& 
	\colhead{${\cal B}_b$}& 
	\colhead{$\sigma({\cal B}_b)$  }& 
	\colhead{${\cal B}_b$ } &
	\colhead{$\sigma({\cal B}_b)$ }
	}
\startdata
491 - 540 & 515 & 2423.0 & 270.8 & - & - & - & - \\ 
 541 - 590 & 565 & 2701.4 & 274.8 & - & - & - & - \\ 
 591 - 640 & 615 & 1952.1 & 205.4 & - & - & - & - \\ 
 641 - 690 & 665 & 1701.9 & 167.7 & - & - & - & - \\ 
 691 - 740 & 715 & 1895.7 & 179.5 & - & - & - & - \\ 
 741 - 790 & 765 & 2386.5 & 202.0 & - & - & - & - \\ 
 791 - 840 & 815 & 2415.5 & 200.5 & - & - & - & - \\ 
 841 - 890 & 865 & 2168.0 & 178.2 & - & - & - & - \\ 
 891 - 940 & 915 & 1567.5 & 138.3 & - & - & - & - \\ 
 941 - 990 & 965 & 1353.6 & 109.9 & - & - & - & - \\ 
 991 - 1040 & 1015 & 976.0 & 89.2 & - & - & - & - \\ 
 1041 - 1090 & 1065 & 1089.5 & 92.9 & - & - & - & - \\ 
 1091 - 1140 & 1115 & 1229.6 & 100.2 & - & - & - & - \\ 
 1141 - 1190 & 1165 & 1169.7 & 88.2 & - & - & - & - \\ 
 1191 - 1240 & 1215 & 861.9 & 73.4 & - & - & - & - \\ 
 1241 - 1290 & 1265 & 819.3 & 68.7 & - & - & - & - \\ 
 1291 - 1340 & 1315 & 740.8 & 60.6 & - & - & - & - \\ 
 1341 - 1390 & 1365 & 746.8 & 60.5 & - & - & - & - \\ 
 1391 - 1440 & 1415 & 782.4 & 64.4 & - & - & - & - \\ 
 1441 - 1490 & 1465 & 718.8 & 55.8 & - & - & - & - \\ 
 1491 - 1540 & 1515 & 619.3 & 50.1 & 559.6 & 43.6 & 585.9 & 87.6 \\ 
 1541 - 1590 & 1565 & 495.6 & 44.5 & 489.0 & 40.2 & 508.0 & 83.0 \\ 
 1591 - 1640 & 1615 & 467.4 & 38.8 & 428.3 & 35.3 & 480.1 & 77.9 \\ 
 1641 - 1690 & 1665 & 380.4 & 34.0 & 382.0 & 32.7 & 446.5 & 74.0 \\ 
 1691 - 1740 & 1715 & 351.6 & 33.6 & 347.8 & 32.1 & 436.8 & 75.0 \\ 
 1741 - 1790 & 1765 & 359.6 & 32.3 & 336.7 & 30.7 & 470.6 & 74.1 \\ 
 1791 - 1840 & 1815 & 334.4 & 29.4 & 336.9 & 28.3 & 396.1 & 65.3 \\ 
 1841 - 1890 & 1865 & 274.8 & 25.3 & 276.6 & 25.8 & 371.5 & 65.0 \\ 
 1891 - 1940 & 1915 & 226.3 & 23.3 & 198.8 & 23.7 & 199.7 & 62.1 \\ 
 1941 - 1990 & 1965 & 259.7 & 23.1 & 254.6 & 23.9 & 310.2 & 61.9 \\ 
 1991 - 2040 & 2015 & 235.8 & 21.6 & 226.0 & 22.5 & 340.2 & 60.6 \\ 
 2041 - 2090 & 2065 & 242.8 & 22.8 & 227.7 & 23.9 & 283.9 & 63.5 \\ 
 2091 - 2140 & 2115 & 168.8 & 18.6 & 206.9 & 21.2 & 268.3 & 57.8 \\ 
 2141 - 2190 & 2165 & 164.7 & 17.8 & 168.4 & 19.7 & 203.4 & 54.9 \\ 
 2191 - 2240 & 2215 & 152.2 & 17.2 & 151.0 & 19.7 & 269.7 & 60.5 \\ 
 2241 - 2290 & 2265 & 132.2 & 15.8 & 114.6 & 18.1 & 206.7 & 56.3 \\ 
 2291 - 2340 & 2315 & 114.9 & 14.7 & 136.5 & 18.1 & 135.8 & 54.2 \\ 
 2341 - 2390 & 2365 & 106.3 & 15.4 & 135.3 & 19.1 & 222.3 & 59.2 \\ 
 2391 - 2700 & 2550 & 82.9 & 5.4 & 99.7 & 6.7 & 157.9 & 21.2 \\ 
 2701 - 3000 & 2850 & 57.5 & 5.3 & 72.6 & 6.5 & 189.1 & 20.8 \\ 
 3001 - 3400 & 3200 & 37.2 & 4.8 & 54.0 & 5.5 & 158.3 & 16.9 \\ 
 3401 - 3800 & 3600 & 39.8 & 5.6 & 50.9 & 5.9 & 145.1 & 16.6 \\ 
 3801 - 4200 & 4000 & 48.1 & 6.8 & 62.9 & 6.7 & 204.5 & 17.9 \\ 
 4201 - 4600 & 4400 & 34.9 & 8.3 & 74.6 & 7.8 & 198.7 & 19.7 \\ 
 4601 - 5000 & 4800 & 36.4 & 10.2 & 98.0 & 9.3 & 248.7 & 22.2 \\ 
 5001 - 5900 & 5450 & 55.8 & 9.6 & 106.3 & 8.3 & 281.7 & 18.9 \\ 
 5901 - 6800 & 6350 & 56.6 & 15.4 & 131.7 & 12.5 & 378.3 & 26.8 \\ 
 6801 - 7700 & 7250 & 127.8 & 25.6 & 127.7 & 19.1 & 463.7 & 38.1 \\ 
 7701 - 8600 & 8150 & 115.3 & 42.4 & 228.0 & 30.0 & 620.6 & 55.6 \\ 
 8601 - 9500 & 9050 & 208.3 & 73.1 & 175.9 & 48.6 & 812.8 & 85.0 \\ 
 9501 - 9900 & 9750 & 316.9 & 172.4 & 152.4 & 108.1 & 695.0 & 176.1 \\ 
 
\enddata
\end{deluxetable*}

\section{DISCUSSION}
\label{sec: conclusions}
We have presented a measurement of the power spectrum of the CMB sky 
observed with the Atacama Cosmology Telescope at \arone\ and \artwo{}. 
The \arone\ spectrum spans a large dynamic range from $\ell=\lbegin$ to 
$\ell=\lend$, covering the damping tail, where the primary anisotropies 
with the higher-order acoustic peaks  dominate,  to the composite high multipole tail
 of the CMB where  emission from dusty galaxies and radio sources and the SZ effect 
 contribute.   The second through the seventh acoustic peaks of the CMB are clearly visible in 
this spectrum. For the \ara$\times$\artwo{} and \arb$\times$ \artwo{} spectra, 
we present measurements from $1500 <\ell <\lend$. The high multipole information 
from these spectra help us constrain the nature of the point source populations that
contribute to this range, and in turn, constrain the SZ contribution to the \arone{}
spectrum.  We measure the presence of a very faint dust signal at high Galactic latitudes.
Recovering such a faint signal gives us confidence in the fidelity of the maps. 
 We find evidence of gravitational lensing of the CMB in the power spectrum at the 2.8$\sigma$
   level. 
Constraints on cosmological parameters and the interpretation of high 
multipole spectra in terms of point source populations and the SZ effect are 
presented in the companion paper by \citet{dunkley/etal:prep}. 
\acknowledgements
 ACT is on the
Chajnantor Science preserve, which was made possible by the Chilean
Comisi\'on Nacional de Investigaci\'on Cient\'ifica y Tecnol\'ogica.
We are grateful for the assistance we received at various times from
the ALMA, APEX, ASTE, CBI/QUIET, and NANTEN2 groups.  
The PWV data come from the public APEX weather website.
Field operations were based at the Don Esteban facility run by
Astro-Norte. Reed Plimpton and David Jacobson
worked at the telescope during the 2008 season. 
We thank Norm Jarosik for support throughout the project. 
This work was supported by the U.S. National Science Foundation
through awards AST-0408698 for the ACT project, and PHY-0355328,
AST-0707731 and PIRE-0507768. Funding was also provided by Princeton
University and the University of Pennsylvania.  The PIRE program made
possible exchanges between Chile, South Africa, Spain and the US that
enabled this research program.  

SD acknowledges support from the Berkeley Center for Cosmological Physics 
Fellowship.  SD would like to thank Christian Reichardt and Oliver Zahn
 for useful discussions. 
 We thank Bruce Bassett for suggestions on testing lensing in the power spectrum.
 SD, AH, and TM were supported through NASA grant NNX08AH30G.  
 JD acknowledges support from an RCUK Fellowship.  RH
received funding from the Rhodes Trust.  
 ADH received additional support from a Natural Science and Engineering
Research Council of Canada (NSERC) PGS-D scholarship. AK and BP were
partially supported through NSF AST-0546035 and AST-0606975,
respectively, for work on ACT\@.  LI acknowledges partial support
from FONDAP Centro de Astrof\'isica.  RD was supported by CONICYT,
MECESUP, and Fundaci\'on Andes.  ES acknowledges support by NSF
Physics Frontier Center grant PHY-0114422 to the Kavli Institute of
Cosmological Physics. KM, M Hilton, and RW received financial support
from the South African National Research Foundation (NRF), the Meraka
Institute via funding for the South African Centre for High
Performance Computing (CHPC), and the South African Square Kilometer
Array (SKA) Project.    YTL acknowledges support from the World Premier
International Research Center Initiative, MEXT, Japan. NS is supported by the 
U.S. Department of Energy contract to SLAC no. DE-AC3-76SF00515. 
We acknowledge the use of the Legacy Archive for
 Microwave Background Data Analysis (LAMBDA). Support for LAMBDA is provided by
  the NASA Office of Space Science.  The data will be made public through LAMBDA (\url{http://lambda.gsfc.nasa.gov/}), 
  and the ACT website (\url{http://www.physics.princeton.edu/act/}).

\appendix 
\section{Cross-frequency Cross-spectrum Errorbars}
Here we derive an analytic expression for the expected error bars on the cross-frequency cross-power spectrum, 
assuming the signal and the noise can be approximated as Gaussian random fields. We do not treat the Poisson 
component from point sources here, as the Poisson noise correction is negligible at our current noise level. 
As before, we denote the two frequency channels as $A$ and $B$ and use lowercase romans $i,j,k,l, etc.$ to 
denote the sub-season data splits. \par
The estimator for the cross-frequency cross-power spectrum in bin $b$ is given by:
\be
\label{eq: cross-spec-mean}
\widehat C^{A\times B}_b = \frac{1}{n_d(n_d-1)} \sum_{i,j; ~i \ne j} \widehat C_b ^{(iA \times jB)},
\ee
where $n_d$ is the number of sub-season data splits (four in our case), and 
\be
\label{eq: cross-spec-indiv}
\widehat C_b^{(iA \times jB)} = \frac{1}{\nu_b}  \sum_{\bl \in b} T^{*iA}_{\bl} T^{jB}_{\bl},
\ee 
where the sum is over pixels in the annulus in Fourier-space contained in the bin $b$, and $\nu_b$ is the effective 
number of  pixels in the annulus over which the sum is taken. Evidently,
\ba
\ave{C_b ^{(iA \times jB)} }  &=& C_b,\\
\ave{\widehat C^{A\times B}_b } &=& C_b.
\ea
First, we will compute the expected covariance between a pair of estimators:
\ba
\nonumber
\Theta^{(iA\times j B);( kA\times l B)}_{bb} = \ave{\left(\widehat C_b ^{(iA \times jB)}-\ave{\widehat C_b ^{(iA \times jB)}}
\right) \left(\widehat C_b ^{(kA \times lB)}-\ave{\widehat C_b ^{(kA \times lB)}}\right)}
\ea
where $i\ne j$ and  $k\ne l$.
From the definition  (\ref{eq: cross-spec-indiv}), we can write
\ba
\Theta^{(iA\times j B);( kA\times l B)}_{bb} = \frac{1}{\nu_b^2}  \sum_{\bl \in b}   \sum_{\bl' \in b} \left[ \ave{T^
{*iA}_{\bl} T^{jB}_{\bl} T^{*kA}_{\bl'} T^{lB}_{\bl'} }\right]  - \ave{C_b ^{(iA \times jB)} } \ave{C_b ^{(kA \times 
lB)} } . 
\ea
Applying Wick's Theorem to the above formula, we get, 
\ba
\nonumber \Theta^{(iA\times j B);( kA\times l B)}_{bb} &=& \frac{1}{\nu_b^2}  \left[ \nu_b^2  \ave{C_b ^{(iA 
\times jB)} } \ave{C_b ^{(kA \times lB)} } + \nu_b \ave{C_b ^{(iA \times kA)} } \ave{C_b ^{(jB \times lB)} } + \nu_b 
\ave{C_b ^{(iA \times lB)} } \ave{C_b ^{(kA \times jB)} }\right ] \\
&&  - \ave{C_b ^{(iA \times jB)} } \ave{C_b ^{(kA \times lB)} } \\
& = &  \frac{1}{\nu_b}  \left[ \ave{C_b ^{(iA \times kA)} } \ave{C_b ^{(jB \times lB)} } + \ave{C_b ^{(iA \times lB)} } 
\ave{C_b ^{(kA \times jB)} }\right] .
\label{eq: correlator}
\ea

Now we will examine various cases of the above variance. \\
{\bf Type 1}: All four splits different, $i\ne j\ne k\ne l$. 
In this case, all four ensemble averages are cross-power spectra and evaluate to the underlying spectrum $C_b$, 
giving
\be
\Theta^{(iA\times j B);( kA\times l B); \text{Type 1}}_{bb}  = \frac{2}{\nu_b} C_b^2. 
\ee
\\
{\bf Type 2a}: One split in common, of channel A, ($i=k\ne l \ne j$ ).
Here, the first ensemble average in  (\ref{eq: correlator}) is an auto-frequency auto-power spectrum which evaluates 
out to $(C_b+N_b^{AA})$, where $N_b^{AA}$ is the noise in the channel A data split. This gives
\be
\Theta^{(iA\times j B);( kA\times l B);  \text{Type 2a}}_{bb}  = \frac{1}{\nu_b} (2 C_b^2+ C_b N_b^{AA}).
\ee 
\\
{\bf Type 2b}: One split in common, of channel B, ($l  = j  \ne k \ne l$ ).
\be
\Theta^{(iA\times j B);( kA\times l B) ; \text{Type 2b}}_{bb}  = \frac{1}{\nu_b} (2 C_b^2+ C_b N_b^{BB}).
\ee 
\\
{\bf Type 2c}: One split in common, one of channel A,  the other of channel B ($i=l$ or $ k= j$ ).
\be
\Theta^{(iA\times j B);( kA\times l B);  \text{Type 2c}}_{bb}  =  \frac{1}{\nu_b} (2 C_b^2+ C_b N_b^{AB})
,\ee
where $N_b^{AB}$ represents any noise that is correlated across the two frequency channels.
\\
{\bf Type 3a}: Two splits in common ($i=k$, $ j= l$ ).
In this case, we have 
\be
\Theta^{(iA\times j B);( kA\times l B); \text{Type 3a}}_{bb}  =  \frac{1}{\nu_b} \left[2 C_b^2+ C_b (N_b^{AA}
+N_b^{BB}) + N_b^{AA} N_b^{BB}\right] .
\ee
\\
{\bf Type 3b}: Two splits in common ($i=l$, $ j= k$ ).
This  evaluates to
\be
\Theta^{(iA\times j B);( kA\times l B); \text{Type 3b}}_{bb}  =  \frac{1}{\nu_b} \left[2 C_b^2+ 2 C_b N_b^{AB} + 
N_b^{AB} N_b^{AB}\right] .
\ee

Next, we turn our attention to the variance of the mean spectrum estimator, $\widehat C_b^{A\times B}$, defined in (\ref{eq: cross-spec-mean}):
\be
\Theta^{\text{mean;mean}} =  \ave{\left(\widehat C^{A\times B}_b -\ave{\widehat C_b^{A\times B}} \right) \left(\widehat C_b^{A\times B} -\ave{\widehat C_b^{A\times B} }\right)} . 
\ee
Expanding one of the terms out, we have
\be
\Theta^{\text{mean;mean}} =  \frac{1}{n_d(n_d-1)} \sum_{i,j; i\ne j }\ave{\left(\widehat C_b^{(iA\times j B)} -\ave{\widehat 
C_b^{(iA\times j B)}} \right) \left(\widehat C_b ^{A\times B} -\ave{\widehat C_b^{A\times B} }\right)} . 
\ee
Each of the $n_d(n_d-1)$ terms in the sum is statistically equivalent to the other and should evaluate to be the 
same. Hence, 
\be
\Theta^{\text{mean;mean}} = \Theta^{(iA\times j B);\text{mean}} ,
\ee
which can be expanded out as
\be
\Theta^{\text{mean;mean}} = \frac{1}{n_d(n_d-1)} \sum_{k,l; k\ne l }\Theta^{(iA\times j B);( kA\times l B)} .
\ee
\par
The last piece of the calculation is to figure out how many times each of the types defined above appear in this sum. 
Since $i$ and $j$ are fixed, it is apparent that for Type 1, $k$ can be chosen in $(n_d-2)$ ways, and $l$ in $(n_d-3)$ 
ways. So, Type 1 appears $(n_d-2)(n_d-3)$ times.  For Type 2a $k$ is fixed by equality with $i$, so $l$ can be chosen 
only in $(n_d-2)$ ways. Therefore, Type 2a appears $(n_d -2)$ times. From symmetry, Type 2b also appears  $(n_d-2)$ 
times. For Type 2c, either $i=l$ or $k=j$ which gives $2(n_d-2)$ terms. Finally, for each of Type 3a and 3b, all four 
indices are completely fixed, so they appear only once. Putting these together, we have:
\ba
\nonumber \Theta^{\text{mean;mean}}  & = & \frac{1}{n_d(n_d-1)} \left [ (n_d-2)(n_d-3)  \frac{2}{\nu_b} C_b^2 
\right .\\
\nonumber & & +  (n_d-2)  \frac{1}{\nu_b} (2 C_b^2+ C_b N_b^{AA})\\
\nonumber & & +  (n_d-2)  \frac{1}{\nu_b} (2 C_b^2+ C_b N_b^{BB})\\
\nonumber & & +  2(n_d-2) \frac{1}{\nu_b} (2 C_b^2+ C_b N_b^{AB})\\
\nonumber & & +  \frac{1}{\nu_b} \left\{2 C_b^2+ C_b (N_b^{AA}+N_b^{BB}) + N_b^{AA} N_b^{BB}\right\}\\
\nonumber & & +  \left .\frac{1}{\nu_b} \left\{2 C_b^2+ 2 C_b N_b^{AB} + N_b^{AB} N_b^{AB}\right \} \right] \\
\nonumber & = & \frac{1}{\nu_b} \frac{1}{n_d(n_d-1)} \left [ n_d(n_d-1) 2 C_b^2 + (n_d-1) C_b (N_b^{AA}+N_b^{BB}) \right .\\
\nonumber && \left. + 2 (n_d-1) C_b N_b^{AB} +  ( N_b^{AA} N_b^{BB} + N_b^{AB} N_b^{AB} ) \right] 
\ea
which simplifies to
\ba
 \Theta^{\text{mean;mean}}  & = &  \frac{1}{\nu_b} \left[ 2 C_b^2 + \frac{C_b}{n_d}  (N_b^{AA}+N_b^{BB}) + 
\frac{2}{n_d} C_b N_b^{AB}+ \frac{(N_b^{AA}N_b^{BB}+N_b^{AB}N_b^{AB})}{n_d (n_d-1)}   \right] .
\ea
Note that if there is only one channel $A\equiv B$,  this reduces to the familiar form \citep[see Eq.~(9) in][]{fowler/etal:prep}:
\ba
 \Theta^{\text{mean;mean}}  & = &  \frac{1}{\nu_b} \left[ 2 C_b^2 + 4 \frac{C_b}{n_d}  N_b^{AA} + 2 \frac{(N_b^{AA})^2}{n_d (n_d-1)}   \right] .
\ea
Finally, if we have $n_p$ patches with equal noise, the above variance should be divided by $n_p$.  In practice, a weighted  combination of the 
individual patch variances is used.  Under the assumption of isotropic noise and filtering, the effect of the data window, $W$, can be taken into account
by correcting the number of modes as $\nu_b \rightarrow \nu_b w_2^2/w_4$, where $w_n$ represents the $n$-th moment of the window.  There is
a small additional 
correction due the anisotropic nature of the noise and  filtering, which we calibrate against Monte Carlo simulations. \
\par

For estimating parameters from the three spectra, $C_b^{\ara \times \ara}$, $C_b^{\ara \times \arb}$ and  $C_b^{\arb \times \arb}$
a joint likelihood function is written as
\begin{equation}
- 2 \ln {\it L} = 
\left(
\begin{array}{c}
 C_b^{\ara \times \ara} - C_b^{\ara \times \ara, \rm{theory}}\\
 C_b^{\ara \times \arb} - C_b^{\ara \times \arb, \rm{theory}}\\
 C_b^{\arb \times \arb} - C_b^{\arb \times \arb, \rm{theory}}
\end{array}
\right)^T
{\bf \Sigma}^{-1}
\left(
\begin{array}{c}
C_b^{\ara \times \ara} - C_b^{\ara \times \ara, \rm{theory}}\\
 C_b^{\ara \times \arb} - C_b^{\ara \times \arb, \rm{theory}}\\
 C_b^{\arb \times \arb} - C_b^{\arb \times \arb, \rm{theory}}
\end{array}
\right),
\end{equation}
where
\begin{equation}
{\bf \Sigma} = 
\left(
\begin{array}{ccc}
\Sigma_b^{\ara-\ara,\ara-\ara} & \Sigma_b^{\ara-\ara,\ara-\arb} & \Sigma_b^{\ara-\ara,\arb-\arb} \\
\Sigma_b^{\ara-\arb,\ara-\ara} & \Sigma_b^{\ara-\arb,\ara-\arb} & \Sigma_b^{\ara-\arb,\arb-\arb} \\
\Sigma_b^{\arb-\arb,\ara-\ara} & \Sigma_b^{\arb-\arb,\ara-\arb} & \Sigma_b^{\arb-\arb,\arb-\arb} 
\end{array}
\right)
\end{equation}
with 
\begin{equation}
\Sigma_b^{A-B,C-D} = \ave{\left(\widehat C^{A\times B}_b -\ave{\widehat C_b^{A \times B}} \right)\left(\widehat C_b^{C\times D} -\ave{\widehat C_b^{C\times D}\right)}},
\end{equation}
and  can be computed in a similar fashion as above.

\end{document}